\documentclass[pra,aps,onecolumn,superscriptaddress,showpacs,nofootinbib]{revtex4-2}

\usepackage[dvips]{graphicx}
\usepackage{amsmath,amssymb,amsthm,mathrsfs,amsfonts,dsfont}
\usepackage{subfigure, epsfig}
\usepackage{physics}
\usepackage{bm}
\usepackage{enumerate}
\usepackage{color}
\usepackage{qcircuit}
\usepackage[normalem]{ulem} % some line form: delete line
\usepackage{epstopdf}
\usepackage{comment}
\usepackage{diagbox}
\usepackage{multirow}    % multiculumn multirow
\usepackage{array}     % column centering
\usepackage{enumitem} % for item without label
\usepackage{booktabs,colortbl} % for long table
\usepackage{rotating}
\usepackage{ulem}
\usepackage{longtable}
\usepackage{setspace}
\usepackage[ruled,vlined]{algorithm2e}
\usepackage[colorlinks = true]{hyperref}
\SetKwInput{KwInput}{Input}
\SetKwInput{KwOutput}{Output}

\graphicspath{{./figure/}}
\renewcommand{\i}{\mathrm{i}}

\begin{document}
\preprint{APS/123-QED}

\title{Experimental Mode-Pairing Measurement-Device-Independent Quantum Key Distribution without Global Phase-Locking}

\author{Hao-Tao Zhu}
\affiliation{Hefei National Research Center for Physical Sciences at the Microscale and School of Physical Sciences, University of Science and Technology of China, Hefei 230026, China}
\affiliation{CAS Center for Excellence in Quantum Information and Quantum Physics, University of Science and Technology of China, Hefei, Anhui 230026, China}
\affiliation{Hefei National Laboratory, University of Science and Technology of China, Hefei, Anhui 230088,China}

\author{Yizhi Huang}
\affiliation{Center for Quantum Information, Institute for Interdisciplinary Information Sciences, Tsinghua University, Beijing 100084, China}

\author{Hui Liu}
\author{Pei Zeng}
\author{Mi Zou}
\affiliation{Hefei National Research Center for Physical Sciences at the Microscale and School of Physical Sciences, University of Science and Technology of China, Hefei 230026, China}
\affiliation{CAS Center for Excellence in Quantum Information and Quantum Physics, University of Science and Technology of China, Hefei, Anhui 230026, China}
\affiliation{Hefei National Laboratory, University of Science and Technology of China, Hefei, Anhui 230088,China}

\author{Yunqi Dai}
\author{Shibiao Tang}
\affiliation{QuantumCTek Corporation Limited, Hefei, Anhui 230088, China}

\author{Hao Li}
\author{Lixing You}
\author{Zhen Wang}
\affiliation{State Key Laboratory of Functional Materials for Informatics, Shanghai Institute of Microsystem and Information Technology, Chinese Academy of Sciences, Shanghai 200050, China}

\author{Yu-Ao Chen} 
\affiliation{Hefei National Research Center for Physical Sciences at the Microscale and School of Physical Sciences, University of Science and Technology of China, Hefei 230026, China}
\affiliation{CAS Center for Excellence in Quantum Information and Quantum Physics, University of Science and Technology of China, Hefei, Anhui 230026, China}
\affiliation{Hefei National Laboratory, University of Science and Technology of China, Hefei, Anhui 230088,China}

\author{Xiongfeng Ma}
\email{xma@tsinghua.edu.cn}
\affiliation{Center for Quantum Information, Institute for Interdisciplinary Information Sciences, Tsinghua University, Beijing 100084, China}

\author{Teng-Yun Chen}
\email{tychen@ustc.edu.cn}
\author{Jian-Wei Pan}
\email{pan@ustc.edu.cn}
\affiliation{Hefei National Research Center for Physical Sciences at the Microscale and School of Physical Sciences, University of Science and Technology of China, Hefei 230026, China}
\affiliation{CAS Center for Excellence in Quantum Information and Quantum Physics, University of Science and Technology of China, Hefei, Anhui 230026, China}
\affiliation{Hefei National Laboratory, University of Science and Technology of China, Hefei, Anhui 230088,China}

%\affiliation[1]{Hefei National Research Center for Physical Sciences at the Microscale and School of Physical Sciences, University of Science and Technology of China, Hefei 230026, China}
%\affiliation[2]{CAS Center for Excellence in Quantum Information and Quantum Physics, University of Science and Technology of China, Hefei, Anhui 230026, China}
%\affiliation[3]{Hefei National Laboratory, University of Science and Technology of China, Hefei, Anhui 230088,China}
%\affiliation[4]{Center for Quantum Information, Institute for Interdisciplinary Information Sciences, Tsinghua University, Beijing 100084, China}
%\affiliation[5]{QuantumCTek Corporation Limited, Hefei, Anhui, China}
%\affiliation[6]{State Key Laboratory of Functional Materials for Informatics, Shanghai Institute of Microsystem and Information Technology, Chinese Academy of Sciences, Shanghai, China}
%
%\affiliation[*]{These authors contributed equally to this work.}

\begin{abstract}
In the past two decades, quantum key distribution networks based on telecom fibers have been implemented on metropolitan and intercity scales. One of the bottlenecks lies in the exponential decay of the key rate with respect to the transmission distance. Recently proposed schemes mainly focus on achieving longer distances by creating a long-arm single-photon interferometer over two communication parties. Despite their advantageous performance over long communication distances, the requirement of phase locking between two remote lasers is technically challenging. By adopting the recently proposed mode-pairing idea, we realize high-performance quantum key distribution without global phase locking. Using two independent off-the-shelf lasers, we show a quadratic key-rate improvement over the conventional measurement-device-independent schemes in the regime of metropolitan and intercity distances.
%\red{We also show that the asymptotic performance of our system can break the linear key rate bound.}
For longer distances, we also boost the key rate performance by 3 orders of magnitude via 304 km commercial fiber and 407 km ultralow-loss fiber.
We expect this ready-to-implement high-performance scheme to be widely used in future intercity quantum communication networks.
\end{abstract}

\maketitle

%\section{Introduction}\label{sc:intro}
Quantum key distribution (QKD)~\cite{bennett1984quantum,ekert1991Quantum}, as a building block of quantum networks, allows remote communication parties to establish a secure key based on the laws of quantum physics~\cite{lo1999Unconditional,shor2000Simple}. Currently, many QKD networks of various sizes have been implemented worldwide, such as metropolitan \cite{peev2009secoqc,sasaki2011field,Tang2016MDInet,Chen2021Implementation} and intercity scales \cite{chen2021integrated}. 
For a metropolitan network, the loss budget between two nodes is around 10 dB \cite{Chen2021Implementation}. Usually, the network users are connected to trusted nodes as service providers. For an intercity network, the single-link loss is typically 20 dB. Often, we need to set up trusted relays outside of cities  \cite{chen2021integrated}. In practice, when one of the trusted nodes is compromised, the network security can be severely damaged \cite{Zhou2022Network}. Also, it is difficult and expensive to ensure the security of relay nodes outside cities. Moreover, due to the complicated construction of single-photon detectors, imperfect detection devices would introduce security loopholes \cite{qi2007time,Lydersen2010Hacking}.

\begin{figure*}[htbp]
\centering \includegraphics[width=16cm]{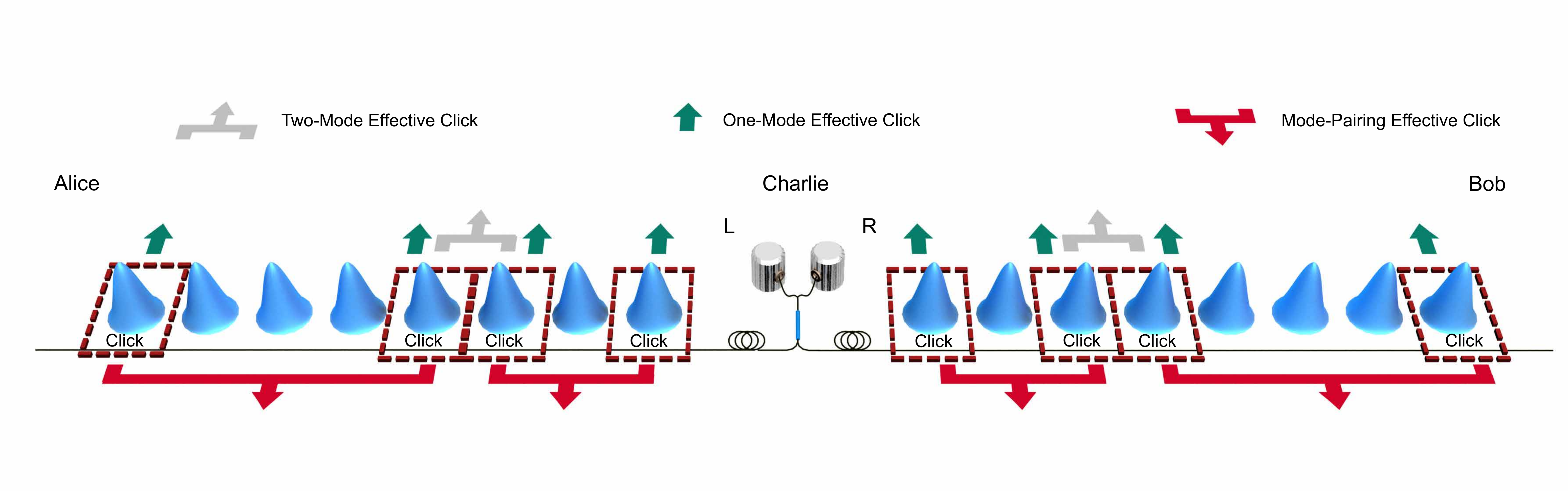}
\caption{Comparison of different MDI QKD schemes. In conventional MDI QKD schemes using two-mode encoding, key information is encoded into two predetermined pulses. Only when Charlie detects both pulses can Alice and Bob learn the encoded information, denoted as gray arrows. In twin-field QKD schemes using one-mode encoding, key information is encoded into one pulse, which is easily disturbed in the channel. When Charlie announces a click on a pulse, denoted as a green arrow, Alice and Bob can derive a key bit. In mode-pairing MDI QKD, Alice and Bob encode information into one pulse to get rid of the coincident detection requirement and pair the clicked pulses based on detection results. They can distill one key bit from any two paired successful detections, denoted as red arrows, which is robust against channel disturbance. 
} \label{fig:MDIQKDdiagram}
\end{figure*}

To close the detection loopholes and reduce the number and cost of trusted nodes, Lo \emph{et al.}~proposed measurement-device-independent quantum key distribution (MDI QKD) \cite{lo2012Measurement}. In a generic MDI QKD setup, as shown in Fig.~\ref{fig:MDIQKDdiagram}, the two communication parties, Alice and Bob, emit encoded laser pulses to a detection site, owned by an untrusted party, Charlie. Charlie employs an interferometer as a quantum relay to correlate the received quantum signals. Charlie announces interference measurement results, based on which Alice and Bob can extract secure key bits. The security of MDI QKD requires no assumption on how Charlie performs measurement and announcement, making it naturally immune to all the detection attacks. Meanwhile, MDI QKD helps reduce the number of trusted nodes and makes quantum communication networks more implementable.

To pursue a practical usage of MDI QKD in metropolitan and intercity quantum networks, we need to consider two main issues: improving the key rate and reducing the experiment requirement. In the conventional MDI QKD schemes, Alice and Bob encode information into two optical modes, such as two adjacent pulses \cite{Ma2012alternative}. This type of encoding, namely two-mode encoding, is relatively simple to implement since it does not require additional devices and modulation. However, the performance of two-mode encoding schemes is limited by the overall channel transmittance $\eta$ since it requires coincidence detection at two optical modes. Another type of MDI QKD, twin-field QKD \cite{lucamarini2018overcoming}, can achieve a quadratic improvement in the key rate. We refer to this as one-mode encoding. In one-mode encoding schemes, Alice and Bob encode information into a single optical pulse, and then Charlie performs a single-photon interference to correlate pulses from two users. This type of encoding, however, necessitates a more challenging experiment implementation because it is more sensitive to environmental noises, especially phase fluctuations of lasers and phase drifts in optical channels. To suppress noises, existing experiments apply advanced global phase-locking techniques~\cite{minder2019experimental,liu2019experimental,chen2020sending,fang2020implementation,Mao2021recent,chen2021twin} to stabilize the phase references between two remote parties, which remains challenging and impractical for large-scale applications.

From the comparison of the existing MDI QKD schemes above, there seems to be a trade-off between high performance and simple implementation. Surprisingly, a recent MDI QKD proposal, mode-pairing quantum key distribution (MP QKD)~\cite{zeng2022quantum}, employs a hybrid encoding method to offer both high performance and simple implementation. See also Ref.~\cite{xie2022breaking} for a similar idea without a rigorous security proof.
Different types of MDI QKD schemes are illustrated in Fig.~\ref{fig:MDIQKDdiagram}.
In MP QKD, Alice and Bob encode key information in a single optical pulse. After Charlie's announcement, they \emph{pair} all the detected locations and generate raw key bits among each pair. 
The core observation of MP QKD is that the two optical modes used to encode the relative information can be determined after Charlie's announcement. 
At the encoding and detecting stage, Alice and Bob only consider a single mode and do not require coincidence detection in predetermined locations. 
At the postprocessing stage, they generate the raw key bits from two pulses and avoid the global phase-locking requirement. Therefore, the users can achieve a quadratic improvement in key rate with simple hardware implementation.

By adopting two off-the-shelf lasers, we realize this high-performance MDI QKD without global phase locking. To this end, we adjust the original MP QKD protocol~\cite{zeng2022quantum} and introduce phase reference estimation techniques to deal with the frequency fluctuation of two independent lasers.
The results show that our implementation can achieve a quadratic key-rate improvement over the conventional MDI QKD schemes.

We now briefly introduce the MP QKD scheme and leave a complete description in Appendix \ref{sc:SchemeDescrip}. In each round, Alice generates a laser pulse of coherent state $\ket{\sqrt{\mu^a}e^{\i \phi^a}}$ with modulated intensity $\mu^a$ randomly chosen from $\{0,\nu,\mu\}$ and modulated phase randomly chosen from $\{0, \frac{2\pi}{D}, \frac{4\pi}{D}, ..., \frac{2\pi(D-1)}{D}\}$.
In this work, we pick up $D=16$ and $0<\nu<\mu<1$. Similarly, Bob generates a coherent state $\ket{\sqrt{\mu^b}e^{\i \phi^b}}$. They then emit the two coherent laser pulses to Charlie, who performs the  interference using a balanced beam splitter and two single-photon detectors $L$ and $R$. Charlie then announces the detection results of each pulse. After repeating the above procedure for many rounds, Alice and Bob postselect all the rounds with successful detection that either $L$ or $R$ clicks. They ``pair'' these rounds and determine the basis and key values on each pair based on the relative intensity and phase information. Afterward, they perform basis sifting, key mapping, and data postprocessing similar to the two-mode MDI QKD schemes.

The final key length of the MP scheme is given by
\begin{equation}\label{eq:keyR}
	K =M^{Z}_{11}  \left[1 - h\left(e^{Z,ph}_{11}\right)\right] -f  M_{\mu\mu} h\left(E_{\mu\mu} \right),
\end{equation}
where $h(x)=-x \log_2 x-(1-x)\log_2(1-x)$ is the binary entropy function and $f$ is the error correction efficiency. The single-photon component of the $Z$-basis pairs, $M_{11}^Z$, and the corresponding phase error rate, $e^{Z,ph}_{11}$, can be estimated by the decoy-state method \cite{hwang2003decoy,Lo2005Decoy,wang2005decoy}. The number of pairs used to distill final key bits, $M_{\mu\mu}$, and the bit error rate, $E_{\mu\mu}$, can be directly obtained from the experiment. For simplicity, here we only use the $Z$-basis pairs with intensities $(\mu,\mu)$ for key generation. The key rate is defined as $R=K/N_{\text{pair}}$, where $N_{\text{pair}}$ is the number of possible pairs and equals half the total number of rounds. 

When we implement MP QKD with two independent lasers, it brings new challenges. When the fiber length increases, the probability of successful detection decreases, enlarging the average pairing length. Consequently, the phase references between the pairs of the two users will drift away due to the phase fluctuation of the lasers and fibers. This leads to a high phase error rate and hence a low key rate. 
To compensate for the phase reference deviation, Alice and Bob need to estimate the underlying phase reference. To do this, for some rounds, they emit strong light pulses without phase modulation for the interference detection. In a 100 \textmu s cycle, Alice and Bob use the first 25.76 \textmu s for strong light pulses, followed by a 3.07 \textmu s recovery region of vacuum state to avoid the cross talks, and the rest 71.17 \textmu s for QKD pulses. After Charlie announces the measurement results, Alice and Bob can use the maximum likelihood estimation (MLE) method to estimate the $\Delta\omega(t)$ with the click results of strong light pulses. 
The likelihood function we use is 
\begin{equation}
f(\Delta\omega)= \sum_{(i,j)} \ln\left\{\dfrac{1}{2}+(-1)^{D_i-D_j}\dfrac{\cos\left[\Delta\omega\tau (j-i)\right]}{4}\right\},
\end{equation}
where $i,j$ denote the locations of the two paired rounds and $D_i,D_j$ denote corresponding results. For convenience, we use $0$ to represent the left detector clicks and $1$ to represent the right one clicks. The time interval between two adjacent pulses is $\tau$. The summation here is taken over all possible pairs of strong light pulses.
Alice and Bob can repeat the estimation steps to get frequency differences for a period of time. They fit the estimated results to obtain $\Delta \omega(t)$ of QKD pulses using 200 periods.  
More details and test results are shown in Appendix \ref{Sc:phaseref}.

\begin{figure*}[htbp]
\includegraphics[width=14cm]{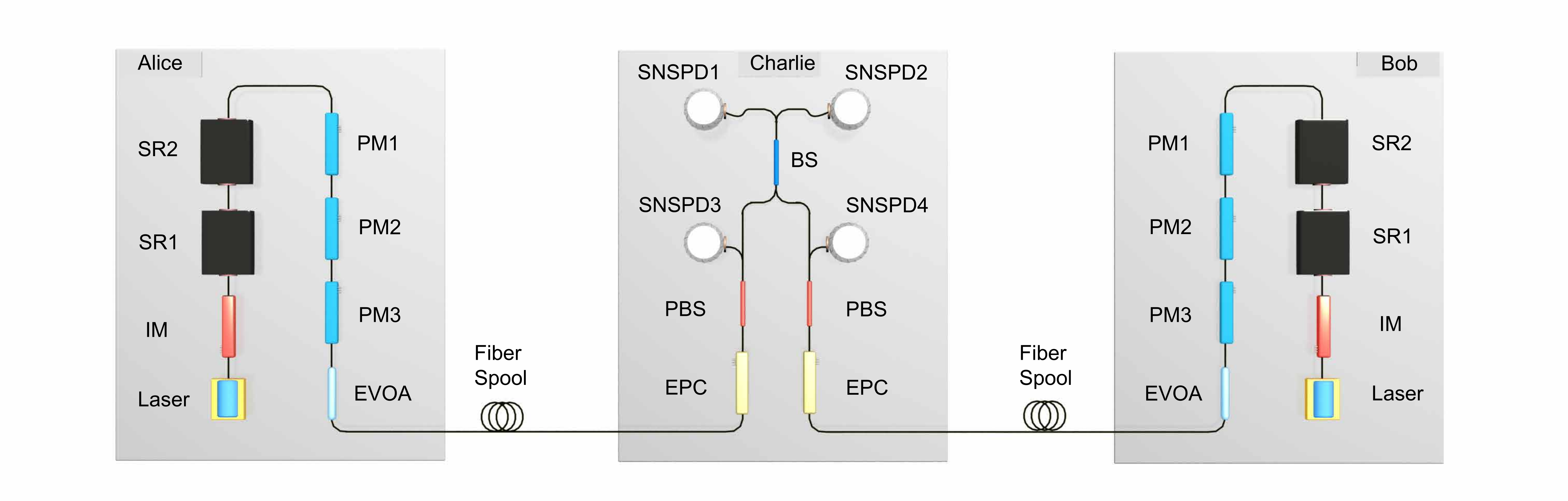}
\caption{Experimental setup. Alice's and Bob's setups are identical, but their encoding modulations are independent. The continuous-wave laser is chopped into discrete pulses by an intensity modulator (IM). Then these pulses are randomly modulated into one of the four intensities---strong, signal, decoy, and vacuum pulses---with the aid of two Sagnac rings (SR1, SR2). Three phase modulators (PM1, PM2, PM3) are used for phase encoding and active phase randomization. The encoded pulses are attenuated to the single-photon level by an electrical variable optical attenuator (EVOA) and transmitted to Charlie. Before interference measurements, the pulse polarisation is aligned by an electric polarization controller (EPC) and a polarization beam splitter (PBS). Finally, the signals are detected by superconducting nanowire single-photon detectors (SNSPDs). SNSPD1 and SNSPD2 are used for interference detection, and SNSPD3 and SNSPD4 are used for polarization feedback and arriving time feedback. Note that we do not carry out any phase-locking operations in the setup.
} \label{fig:setup}
\end{figure*}

The experimental setup is shown in Fig.~\ref{fig:setup}. Alice and Bob employ the off-the-shelf continuous-wave lasers (ORION 1550 nm Laser Module) whose linewidth is 2 kHz and center wavelength is 1550.12 nm. An intensity modulator chops the emitted light into pulses of width 400 ps at 625 MHz. Then, the key and basis information is encoded into these pulses by two Sagnac rings and three phase modulators for different intensities and phases. Afterward, pulses are attenuated to the single-photon level by an electrical variable optical attenuator and transmitted to Charlie for interference detection. More details for the setup are shown in Appendix \ref{sc:ExprDetail}.

We consider the experimental settings under the scenario of metropolitan and intercity quantum networks. 
For a metropolitan (intercity) network, the loss budget between Alice to the measurement site is around 10 dB \cite{Chen2021Implementation} (20 dB \cite{chen2021integrated}), corresponding to 100 km (200 km) fiber from Alice to Bob when using a symmetric channel. 
A longer intercity communication distance of 300 to 400 km is also of practical interest.
Hence, we perform the experiment via 101, 202, 304 km standard and 407 km ultralow-loss optical fibers. The main experiment parameters are listed in Table~\ref{table:ExpParametersII}.

\begin{table}[bpth!]
%\vspace{10pt}
\centering
\caption{Experimental parameters. The mean photon numbers of signal and decoy states are denoted as $\mu$ and $\nu$, respectively. The total transmittance of a single side is $\eta$. The total pulses sent by users is $N$. The maximum pairing length is $L_{\text{max}}$. The detector dark count is about 34 Hz, corresponding to a rate of 2.72$\times 10^{-8}$/pulse. The detection efficiency is $\eta_d=62.46\%$ including the insertion loss of 0.58 dB. The error-correction efficiency is $f=1.1$. The security parameter is $\epsilon=10^{-10}$.}
\label{table:ExpParametersII}
\begin{tabular}{l|llll}
	\hline
	\hline
 & $101$ km  & $202$ km & $304$ km & $407$ km \\  
	\hline		
	$\mu$ & $0.309$ & $0.338$ & $0.531$ & $0.429$ \\
	$\nu$ & $0.032$ & $0.035$ & $0.053$ & $0.038$ \\
	%$\alpha$ & $ 0.208 $ & $ 0.181 $ & $ 0.186 $  & $ 0.162 $ \\
	$\eta$ &$4.32\times10^{-2}$&$6.80\times10^{-3}$&$7.43\times10^{-4}$&$2.18\times10^{-4}$\\
	$\emph{N}$ & $ 5.07\times10^{11} $ & $ 2.10\times10^{12} $ & $ 6.33\times10^{12} $  & $ 7.66\times10^{13} $ \\
	$L_{\max}$& 500& 1000 & 2000 & 2000 \\
	\hline    
	\hline    	
\end{tabular}

\end{table}

After Alice and Bob compensate for the phase differences using estimated $\Delta\omega(t)$, they can use the $X$-basis error rate to quantify how well they have estimated the phase reference. We test the $X$-basis error rate and the number of pairs under different pairing lengths and communication distances. The results show there is a trade-off: a larger $l$ results in more pairs but increases the $X$-basis error rate. In practice, Alice and Bob can set a proper maximum pairing length $L_{\max}$, beyond which they do not pair the corresponding clicks. For a short distance (101 km), the successful detection probability is high and the average pairing length is small, so we pick up $L_{\max}=500$. The average pairing length is larger for a longer distance (202 km), so we pick up $L_{\max}=1000$. For the cases of 304 km and 407 km, we pick up $L_{\max}=2000$, considering pairs with $l>2000$ have a relatively high $X$-basis error rate and contribute little to the final key rate. We give the detailed results in Appendix \ref{Sc:detailest}.

\begin{figure}[htbp]
	\includegraphics[width=12cm]{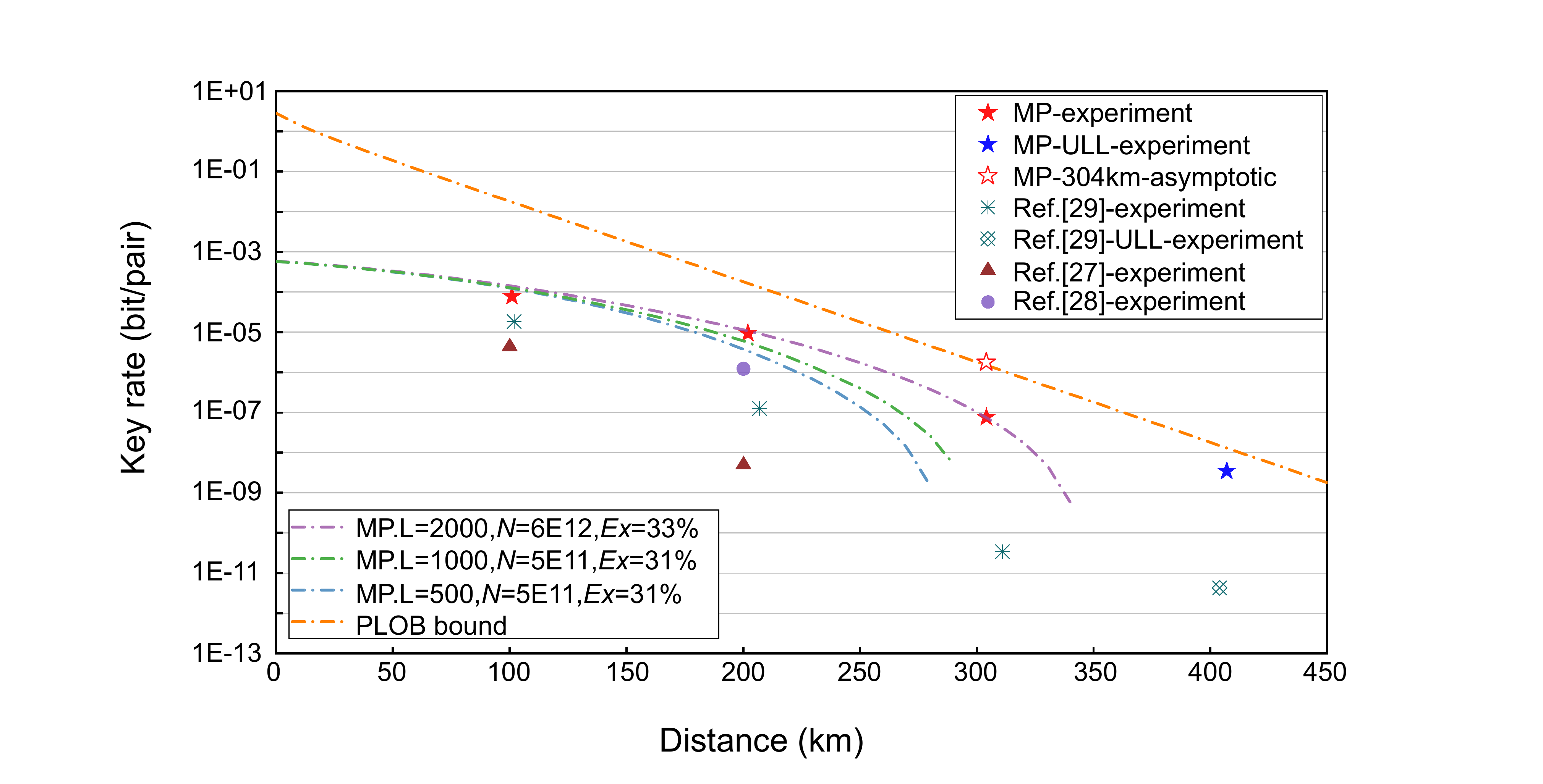}
	\caption{Key-rate performance. The experimental rate–distance performance of MP QKD, compared with the theoretical simulations, along with the existing two-mode MDI QKD experimental results~\cite{semenenko2020chip,woodward2021gigahertz,yin2016measurement} and the linear key rate bound \cite{pirandola2017fundamental}. 
	Data points marked by red and blue stars are key rates of our system using commercial fibers and ultralow loss (ULL) fibers, respectively. We also calculate the asymptotic key rate for the 304 km case based on the experimental data, marked by a hollow red star. Here, $N$ is the total number of QKD rounds, $L$ is the maximum pairing length, and $E_X$ is the $X$-basis error rate. The theoretical expectation curves are simulated under different maximum pairing lengths, data sizes, and $X$-basis error rates. 
	}
	\label{fig:keyrate}
\end{figure}

The key rates for different transmission distances are presented in Fig.~\ref{fig:keyrate}. Here, the $Z$-basis error rate is in the order of $10^{-4}$ with the two Sagnac rings and the intensity modulator, giving over 40 dB of extinction ratio for the signal and vacuum states. We also compare the experimental results with numerical simulations along with previous
experiments. As shown in the key-rate figure, the ratio between the key rate and the square root of the transmittance is given by $\frac{7.75\times10^{-5}}{4.32\times10^{-2}}=1.80\times10^{-3}$ for 101 km and the ratio is $\frac{9.34\times10^{-6}}{6.80\times10^{-3}}=1.37\times10^{-3}$ for 202 km. The results show that under the intercity communication distances (101 and 202 km), the key rate-transmittance relation of our system follows $R=O(\sqrt{\eta})$ rather than $O(\eta)$, indicating a quadratic improvement in the key rate. 

For longer communication distances, even with higher $X$-basis error rates caused by larger phase fluctuations, the system can still maintain a key rate-transmittance relationship well above $R=O(\eta)$. Our system realizes key rates of 19.2 and 0.769 bits per second, respectively, via 304 km and 407 km fibers, 3 orders of magnitude higher than those of the existing MDI QKD experiments \cite{yin2016measurement}. Besides, we calculate the asymptotic key rate of the 304 km case with the experimental data and show that our system has the potential to break the linear key rate bound \cite{pirandola2017fundamental}.
For the one-mode MDI QKD schemes, the key rate is zero under the same setting since the phase differences between Alice and Bob are almost random without global phase locking. We give more comparison and discussion in Appendix \ref{sc:discussion}.

Our experiment shows that the MP QKD scheme owns clear advantages over the existing MDI QKD implementations, especially in the regime of metropolitan and intercity distances.  
We anticipate the MP QKD system and similar designs to improve the performance of quantum communication networks. Also, we expect that the design of the MP QKD experiment will be helpful for the construction of quantum repeaters \cite{Zukowski1993Event,Briegel1998Repeaters}, as well as extending the reach of the quantum internet.

In the future, the scheme has a few potential directions to explore. 
First, in terms of instrument hardware, increasing the system repetition rate is more beneficial to improving the key rate of the MP QKD scheme compared with other MDI QKD schemes. 
A higher repetition rate leads to shorter time intervals between the pulses. 
Hence, the users can choose a larger pairing interval to obtain more pairs. In addition, this is advantageous to the phase reference estimation and results in a lower $X$-basis error rate. The users can also use frequency multiplexing to achieve a similar goal since one can pair detection events from different spectrum channels in the mode-pairing scheme. Other improvements, such as narrowing the linewidth, or improving the stability, will also help increase the key rate further. Theoretically, applying more efficient pairing strategies could also benefit the key rate. Because of the low click probability caused by the low intensity and phase matching of the $X$-basis pairs, the number of them is relatively small, which affects the accuracy of parameter estimation. One possible solution is that re-pairing the clicks that fail to pair during key mapping stage. Intuitively, the pairing process does not reveal anything about the final key bits, so repairing $X$-basis pairs should not affect security.

\acknowledgements
The authors acknowledge Y.~Yan and G.~Liu for the insightful discussions and B.~Bai and J.~Zhang for providing assistance on electronics. This work has been supported by the National Key R\&D Program of China (Grants No.~2017YFA0303903 and No.~2019QY0702), the Chinese Academy of Science, the National Fundamental Research Program, the National Natural Science Foundation of China (Grants No.~11875173, No.~61875182 and No.~12174216), Anhui Initiative in Quantum Information Technologies, and Fundamental Research Funds for the Central Universities (WK2340000083).

H.-T.~Z. and Y.H. contributed equally to this work.

\appendix
\section{Mode-pairing Scheme} \label{sc:SchemeDescrip}
In this section, we first introduce the experiment steps of the mode-pairing (MP) scheme with minor adjustment on phase reference compensation. Then, we present the pairing strategy and data postprocessing. Finally, we show the maximum likelihood estimation (MLE) method for the phase reference estimation.

\subsection{MP QKD scheme with decoy state}\label{sec:protocol}
Below, we present the steps of the mode-pairing quantum key distribution (MP QKD) scheme.
\begin{enumerate}
	\item
	\textbf{State preparation:} Alice and Bob divide the optical pulses into two categories: quantum key distribution (QKD) pulses and strong light pulses. For the $i$-th QKD pulse $(i = 1, 2, ...,N)$, Alice prepares coherent state $\ket{\sqrt{\mu_i^a}\exp(\i \phi^a_i)}$ on the
	optical mode $A_i$ with intensity $\mu_i^a$ chosen from $\{0,\nu,\mu\}$ ($0<\nu<\mu<1$) randomly and phase $\phi^a_i$ uniformly
	chosen from $\{0, \frac{2\pi}{D}, \frac{4\pi}{D}, \cdots, \frac{2\pi(D-1)}{D}\}$. In this work, we pick up $D=16$ and $0<\nu<\mu<1$. Similarly, Bob randomly chooses $\mu_i^b,\phi^b_i$ and prepares $\ket{\sqrt{\mu_i^b}\exp(\i \phi^b_i)}$ on mode $B_i$. For strong light pulses, Alice and Bob prepare pulses on $A_i$ and $B_i$, respectively, without any phase modulation. They choose strong intensities so that they can obtain enough detection clicks on Charlie's side.
	
	\item
	\textbf{State transmission and measurement:} Alice and Bob send the pulses on modes $A_i$ and $B_i$, respectively, to Charlie in the following order: $N_s$ strong light pulses, followed by $N_v$ vacuum pulses as a recovery region and $N_q$ QKD pulses, and then repeat the sequence. More details are shown in Sec.~\ref{Sc:Phase drift}. Charlie is supposed to perform the single-photon interference measurement and announces the clicks of detectors $L$ and/or $R$.
	\item
	Alice and Bob repeat the above two steps until each of them sends $N$ QKD pulses. Then, they perform the following data postprocessing procedures.
	
	\item
	\textbf{Mode pairing:} For all rounds with successful detection ($L$- or $R$-click) of QKD pulses, Alice and Bob use the strategy introduced in Sec.~\ref{subsec:pairstrategy} to group two clicked rounds as a pair. 
	
	\item
	\textbf{Basis assignment:} Alice and Bob label all the pairs of QKD pulses, as data pairs, according to the intensities of the two rounds using Table~\ref{table:SingleSideBasis}. For the paired $i$-th and $j$-th rounds, the corresponding intensities are $\mu^{a(b)}_i$ and $\mu^{a(b)}_j$, respectively.
	\begin{table}[hbpt!] 
		\caption{Basis assignment depending on intensities.}\label{table:SingleSideBasis}
		\begin{tabular}{|c|ccc|}  
			\hline   
			\diagbox{$\mu^{a(b)}_i$}{$\mu^{a(b)}_j$} & 0 & $\nu$&$\mu$ \\   
			\hline   
			0		& `0' 	& $Z$	&$Z$  \\ 
			$\nu$ 	& $Z$ 	& $X$  	&`discard'\\  
			$\mu$	& $Z$ 	& `discard'& $X$  \\      
			\hline   
		\end{tabular} 
	\end{table}
	
	\item \textbf{Basis sifting:}
	Alice and Bob announce the basis of all the data pairs and the sum of the intensities $\vec{\mu}\equiv(\mu_i^a+\mu_j^a,\mu_i^b+\mu_j^b)$ of each pair. Then, they further sift the basis of this data pair according to Table~\ref{table:TwoPartyBasis}. %Note that Alice and Bob will not take use of the data pairs labeled as ``discard'' in the following steps. Finally, they label the data pairs as $Z$-pairs and $X$-pairs accordingly.
	\begin{table}[hbpt!]
		\caption{Basis sifting according to intensity information from the two sides.}\label{table:TwoPartyBasis}
		\begin{tabular}{|c|cccc|}   
			\hline   
			\diagbox{Alice}{Bob} & `0' & $Z$&$X$&`discard' \\   
			\hline   
			0		& `0' 	& $Z$-pair	&$X$-pair  &`discard'\\ 
			$Z$ 	& $Z$-pair 	& $Z$-pair  	&`discard'&`discard'\\  
			$X$		& $X$-pair 	& `discard'& $X$-pair  &`discard'\\      
			`discard'&`discard'&`discard'&`discard'&`discard'\\
			\hline   
		\end{tabular} 
	\end{table}
	
	\item
	\textbf{Phase estimation:} Alice and Bob repeatedly pair the strong light pulses with a successful detection belonging to the same $N_s$ group to obtain all possible strong light pairs under the restriction of maximum and minimum pairing length $L_{max}$ and $L_{min}$. Then they use the MLE method and fitting the results to estimate background phase reference $\Delta\omega(t)$ between the pulses emitted by Alice and Bob. Details of phase estimation are introduced in Sec.~\ref{subsec:MLE}
	
	\item
	\textbf{Key mapping:} For each $Z$-pair on location $i$, $j$, Alice sets her key to $\chi_a=0$ if the intensity of the $i$-th pulse is $\mu_i^a=0$ and $\chi_a=1$ if $\mu_j^a=0$. For each $X$-pair on location $i$, $j$, the key is extracted	from the relative phase, $\chi_a=\lfloor(\phi^a_j-\phi^a_i)/\pi \mod 2\rfloor$ and Alice announces $\theta_a=(\phi^a_j-\phi^a_i)\mod \pi$. Bob also assigns his raw key bits $\chi_b$ and announces $\theta_b$. The only difference is that, for $Z$-pairs Bob sets the raw key bit $\chi_b$ to
	be $1$ if $\mu_i^b=0$ and $\chi_b=0$ if $\mu_j^b=0$.
	For the $X$-pairs, if $|\theta_b-\theta_a| \in \left[ \Delta\theta_{i,j}+2k\pi-\frac{\pi}{D},\Delta\theta_{i,j}+2k\pi+\frac{\pi}{D}\right]$, where $k$ is an integer, $\Delta\theta_{i,j} = \int_{t_i}^{t_j}\Delta\omega(t) dt$, and ${t_i},{t_j}$ are the corresponding time of the $i$-th and $j$-th rounds, Alice and Bob keep the key; otherwise, they discard it.
	
	\item
	\textbf{Parameter estimation:} Alice and Bob use $Z$-pairs with different intensity settings to estimate the number of clicked single-photon pairs $M^Z_{11}$ using the decoy-state method. The $X$-pairs are used to estimate the single-photon phase-error rate $e^{Z,ph}_{11}$ \cite{zeng2022quantum}. The Chernoff-Hoeffding method \cite{Zhang2017improved} is also used to deal with statistical fluctuation. More details are shown in Sec.~\ref{sc:finitesize}. Note that we only use $X$-pairs with intensity $\vec{\mu}=(2\nu,2\nu)$ for parameter estimation here. %in practice for the accuracy.%Note that considering the data size, we only use $X$-pairs with intensity $\vec{\mu}=(2\mu,2\mu)$ for parameter estimation in practice.
	
	\item
	\textbf{Key distillation:} Alice and Bob use the $(\mu,\mu)$-pairs to generate key bits. They perform error correction and privacy amplification according to the key rate formula evaluated by $M^Z_{11}$, $e^{Z,ph}_{11}$, $M_{\mu \mu}$ and $E_{\mu \mu}$. The key rate after considering finite-size effect is given by,
	\begin{equation} \label{eq:keylength}
		M_R=M^{Z,L}_{11}  \left[1 - h\left(e^{Z,ph,U}_{11}\right)\right] -f  M_{\mu\mu} h\left(E_{\mu\mu} \right),
	\end{equation}
	where $M^{Z,L}_{11}$is the lower bound of the single-photon component of the $Z$-basis pairs and $e^{Z,ph,U}_{11}$ is the upper bound of the corresponding phase error rate. They can be estimated by the decoy-state method and the Chernoff-Hoeffding bound.
\end{enumerate}

\subsection{Pairing and key mapping}\label{subsec:pairstrategy}
Here, we take Alice's side as an example to introduce how to pair QKD rounds with successful detection and map the corresponding key bits. We denote Charlie’s announcement in $i$-th round as a binary variable $C_i$: $C_i=1$ for the case that either detector $L$ or $R$ clicks and $C_i = 0$ for the case that no detector clicks or both click.
%\pei{find a way to split the following algorithm!}

\begin{centering}
	\begin{algorithm}[H]\label{algo1}
		\SetAlgoLined
		\LinesNumbered
		\KwInput{The number of QKD rounds $N$; the encoded intensity and phase of each round $\mu_i$ and $\phi_i$, respectively, for $i=1$ to $N$; Charlie's announced detection results $C_i$, for $i=1$ to $N$; maximum and minimum pairing length $L_{max}$ and $L_{min}$.}
		\KwOutput{Front- and rear-pulse locations $(F_j,R_j)$, the basis information $B_j$ and the key bit $K_j$ for the $k$-th pair, for $j=1$ to $N_{pair}$, where $N_{pair}$ is the number of pairs.}
		Initialise the pairing index $j:= 1$; initialise the starting point $i:=1$; initialise the flag $flag:=0$\;
		\textbf{while }{$i<=N$}\textbf{ do}\\
		\quad\textbf{while }{($i<=N$) and ($C_i=0$)}\textbf{ do}\\
		\quad\quad$i:=i+1$;\\
		\quad\textbf{end while}\\
		\quad\textbf{if }{flag=0}\textbf{ then}\\
		\quad\quad set the flag to $flag:=1$; set the front location of $j$-th pair to $F_j:=i$;\\
		\quad\textbf{else}\\
		\quad\quad set the rare location of $j$-th pair to $R_j:=i$;\\
		\quad\quad calculate the pair length of current pair, $l:=R_j-F_j$;\\
		\quad\quad \textbf{if }{($l<=L_{max}$) and ($l>=L_{min}$)}\textbf{ then}\\
		\quad\quad\quad	the $j$-th pair is successfully paired, set the flag to $flag:=0$, update the pair count $j:=j+1$;\\
		\quad\quad\quad \textbf{if }{$(\mu_{F_j},\mu_{R_j})\in\{(\mu,0),(0,\mu),(\nu,0),(0,\nu)\}$}\textbf{ then}\\
		\quad\quad\quad\quad current pair is $Z$-pair, set $B_j:=Z$;\\
		\quad\quad\quad\quad set the key bit:\\
		\quad\quad\quad\quad \textbf{if }{$\mu_{F_j}=0$}\textbf{ then}\\
		\quad\quad\quad\quad \quad$k_j:= 0$;\\
		\quad\quad\quad\quad\textbf{else} \\
		\quad\quad\quad\quad\quad$k_j:= 1$;\\
		\quad\quad\quad \textbf{else if }{$(\mu_{F_j},\mu_{R_j})\in\{(\mu,\mu)(\nu,\nu)\}$}\textbf{ then}\\
		\quad\quad\quad\quad current pair is $X$-pair, set $B_j:=X$;\\
		\quad\quad\quad\quad set the key bit $k_j = \lfloor|\phi_{F_j}-\phi_{R_j}|/\pi\rfloor$;\\
		\quad\quad\quad\quad set the relative phase $\theta_j:=|\phi_{F_j}-\phi_{R_j}|$ mod $\pi$;\\
		\quad\quad\quad\textbf{end if}\\
		\quad\quad\textbf{else}\\
		\quad\quad\quad the pair of $(F_j,R_j)$ fails, update $F_j:=R_j$;\\
		\quad\quad\textbf{end if}\\
		\quad\textbf{end if}\\
		\textbf{end while}\\
		set the number of pairs $j:=j-1$.
		\caption{Pairing and key mapping}
	\end{algorithm}
\end{centering}

\subsection{Finite-size analysis method}\label{sc:finitesize}
To deal with the data fluctuation, we introduce the finite-data-size analysis method. Here, we use the Chernoff-Hoeffding method and follow the conclusion in \cite{Zhang2017improved}. Given an observed quantity $\chi$, the upper and lower bounds of the underlying expectation value is given by,
\begin{equation}
	\begin{split}
		\mathbb{E}^L(\chi)&=\dfrac{\chi}{1+\delta^L},\\
		\mathbb{E}^U(\chi)&=\dfrac{\chi}{1-\delta^U},
	\end{split}
\end{equation}
where $\delta^{L(U)}$ is the solution of
\begin{equation}\label{eq:deltasolve}
	\begin{split}
		\left[\dfrac{e^{\delta^L}}{(1+\delta^L)^{1+\delta^L}}\right]^{\chi / (1+\delta^L)}&=\dfrac{1}{2}\varepsilon,\\
		\left[\dfrac{e^{-\delta^U}}{(1-\delta^U)^{1-\delta^U}}\right]^{\chi / (1-\delta^U)}&=\dfrac{1}{2}\varepsilon.
	\end{split}
\end{equation}
Here $\varepsilon$ is the security parameter whose value is $10^{-10}$ as given in the main text.

\section{Phase reference estimation}\label{Sc:phaseref}
In this section, we first analyze the causes of phase reference differences. Then we introduce the maximum likelihood estimation method we use to estimate phase reference differences. Finally, we show some results obtained by this method

\subsection{Analysis of phase reference differences} \label{Sc:Phase drift}
The phase difference between two pulses from Alice and Bob is mainly caused by two parts: laser and fiber fluctuations. The phase difference between two pulses is given by
\begin{equation} \label{eq:phasediff}
	\theta_{ba} = \theta^{0}_{ba} + (\theta^{fiber}_b - \theta^{fiber}_a) + (\omega_b - \omega_a) t,
\end{equation}
where $ \omega_a $ and $ \omega_b $ are angular frequencies of light pulses and $ t $ is the transmitting time. The transmitting time of Alice's and Bob's pulses is the same with the system synchronization. The additional phase drifts after the pulses going through the long optical fibers are denoted as $\theta_a^{fiber} $ and $ \theta_b^{fiber} $. The initial phase difference between Alice's and Bob's lasers is $ \theta^{0}_{ba} $.

For phase-matching QKD, the first term of Eq.~\eqref{eq:phasediff} is a constant with the phase locking technology. Meanwhile, the second item of the equation can be compensated or calculated as well. However, the first term will vary rapidly without global phase locking and the $X$-basis error rate will be nearly 50 \%, resulting in no secure key.

In MP QKD, Alice and Bob mainly care about the difference of phase differences of the paired $i$th and $j$th rounds, 
\begin{equation}\label{eq:MPphasediff}
	\begin{split}
		\Delta\theta_{i,j} &= (\theta^{0}_{ba,j} - \theta^{0}_{ba,i}) + (\theta^{fiber}_{b,j} - \theta^{fiber}_{a,j}) - (\theta^{fiber}_{b,i} - \theta^{fiber}_{a,i}) + (\omega_b-\omega_a)(t_j-t_i)\\
		&= \Delta\theta^{0}_{ba} + (\Delta\theta^{fiber}_b - \Delta\theta^{fiber}_a)+\Delta \omega (t_j-t_i),
	\end{split}
\end{equation}
where $t_i$ and $t_j$ are the sending time of $i$th and $j$th rounds, respectively. The additional phase differences caused by fibers between the $i$th and $j$th rounds from each side are denoted by $\Delta\theta^{fiber}_a$ and $ \Delta\theta^{fiber}_b $. The difference of initial phase differences is denoted as $\Delta\theta^{0}_{ba}$. 
Equation~\eqref{eq:MPphasediff} shows three main factors contributing to the phase reference differences: %the first term comes from the initial global phase drift due to the limited 2 kHz laser linewidth, the second one is the phase drift caused by fibers, and the third one is the frequency difference between the two unlocked lasers.

\begin{enumerate}
	\item 
	For the first term caused by the limited 2 kHz laser linewidth, the maximum pairing length $L_{\text{max}}$ in the experiment is at most 2000, corresponding to 3.2 \textmu s at our 625 MHz system frequency. So, the phase reference caused by the linewidth is less than $3.2 \text{ \textmu s} \times 2 \text{ kHz} = 6 \times 10^{-3} \text{ rad}$, which is negligible. Besides, the phase difference of this part will further decrease with the increase of system frequency and will not increase with the communication distance.
	\item %Phase drift caused by fiber link
	For the second term caused by fiber link, Ref.~\cite{fang2020implementation} shows that the phase drift in fibers is less than 20 rad/ms. The reference phase drift caused by fiber can be estimated by $l\times 1.6\text{ ns} \times 20 \text{ rad/ms}$, which will increase as the pairing length $l$ increases. Similarly, since $L_{\text{max}}$ is at most 2000 in the MP scheme, the maximum reference phase drift caused by fiber is $l\times 1.6\text{ ns} \times 20 \text{ rad/ms}= 0.06 \text{ rad}$ , which is also negligible. Besides, the phase drift of deployed commercial fiber could be smaller than lab fiber spools \cite{liu2021field}, so the fiber phase drift influence will be smaller than the current results.
	\item In our setup, we mainly need to consider the third term in Eq.~\eqref{eq:MPphasediff}, which is the frequency difference between two remote unlocked lasers. Considering that the angular frequency difference $\Delta \omega$ varies over time, the third term can be approximated as,
	\begin{equation}\label{eq:MPfreq}
		\Delta\theta_{i,j} = \int_{t_i}^{t_j}\Delta\omega(t) dt,	
	\end{equation}
	which is used in the key mapping step of Sec.~\ref{sc:SchemeDescrip}.
\end{enumerate}

\subsection{Maximum-likelihood estimation method for phase reference estimation}\label{subsec:MLE}
When we implement MP QKD with two independent off-the-shelf lasers, it also brings new experimental challenges. When the fiber length increases, the probability of successful detection will decrease, which enlarges the average pairing length. As a consequence, the phase references between the pairs of the two users will drift away due to the phase fluctuation of the two independent lasers and fibers. This leads to a high phase error rate and hence a low key rate. 

\begin{figure*}[htbp]
	\includegraphics[width=12cm]{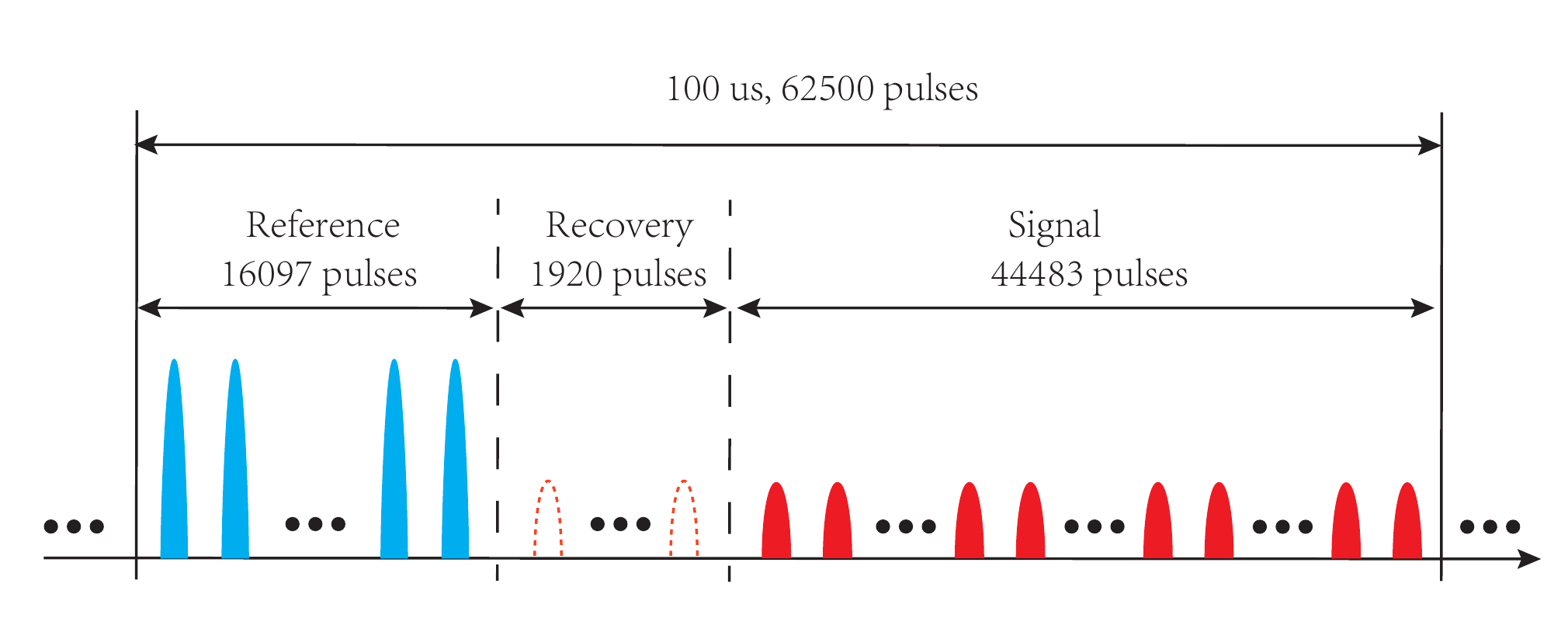}
	\caption{Pulse sequence in one cycle. Alice and Bob send pulses periodically where one cycle is 100 \textmu s. In the reference region, Alice and Bob do not carry out phase modulation on the strong light pulses. In the recovery region, Alice and Bob modulate pulses into vacuum states to avoid the influence caused by strong light pulses. In the signal region, Alice and Bob send pulses with different intensities and phases. After interference, Charlie announces the clicks of all detectors.}
	\label{fig:SM_timesquence}
\end{figure*}

In order to compensate the phase reference deviation, Alice and Bob need to estimate the underlying phase reference. 
As mentioned in the protocol description, to calculate the phase reference between Alice’s and Bob’s lasers, they send strong light pulses periodically. That is, Alice and Bob divide one period into three regions: reference, recovery and signal, as shown in Fig.~\ref{fig:SM_timesquence}. During the recovery region, Alice and Bob send vacuum states and Charlie does not do any measurement or announcement. The recovery region is mainly used to separate the other two regions to prevent the after-pulses of strong pulses from affecting quantum signals. 

In phase-estimation step, they divide the click results of strong light pulses into different data groups and apply the MLE method to each group. In practice, the time interval between the first and last clicks in the same data group is no more than 1 ms so that $\Delta\omega(t)$ can be treated as a constant within a data group. For simplicity, we omit the variable $t$ when estimating the phase reference using one data group.

For click data in a group, Alice and Bob pair arbitrary two click events. For example, they pair $i$th and $j$th clicks. For each click, there are two results, $L$-click (0) or $R$-click (1). Denote the $i$th and $j$th click results as $D_i$ and $D_j$, respectively. Then, we can calculate the probabilities of the same detector clicks twice, $(0,0)$ or $(1,1)$, $P^0$, and the two detectors each clicks once, $(0,1)$ or $(1,0)$, $P^1$,
\begin{equation}
	\begin{aligned}
		P^0 &= \dfrac{1}{2}+\dfrac{\cos\Delta\theta}{4},\\
		P^1 &= \dfrac{1}{2}-\dfrac{\cos\Delta\theta}{4}.\\
	\end{aligned}
\end{equation}
Here, the phase difference between the two pulses, $\Delta\theta$, can be written as
\begin{equation}
	\Delta\theta = \Delta\omega\tau (j-i),
\end{equation}
where $\tau$ is the time interval between two adjacent pulses.
Thus, the likelihood function is given by
\begin{equation}
	\begin{aligned}
		f(\Delta\omega) &=\ln \prod_{(i,j)} P^{|D_i-D_j|} \\
		&= \sum_{(i,j)} \ln\left\{\dfrac{1}{2}+(-1)^{D_i-D_j}\dfrac{\cos\left[\Delta\omega\tau (j-i)\right]}{4}\right\}. \\
	\end{aligned}
\end{equation}
The value of $\Delta\omega$ can be estimated by looking for the maximum of the likelihood function within a predetermined range from preliminary system tests. For example, the range can be read from instruments such as a frequency counter.

Alice and Bob can repeat the above steps to get frequency differences for each data group. They fit the estimated results to obtain $\Delta \omega(t)$ of QKD pulses using 200 data groups. In reality, $\Delta \omega(t)$ changes slowly over time, as shown in Sec.~\ref{Sc:detailest}, so Alice and Bob use the least square method for fitting.

\subsection{Frequency difference estimation results}\label{Sc:detailest}

\begin{figure*}[htbp!]
	\centering\includegraphics[width=16cm]{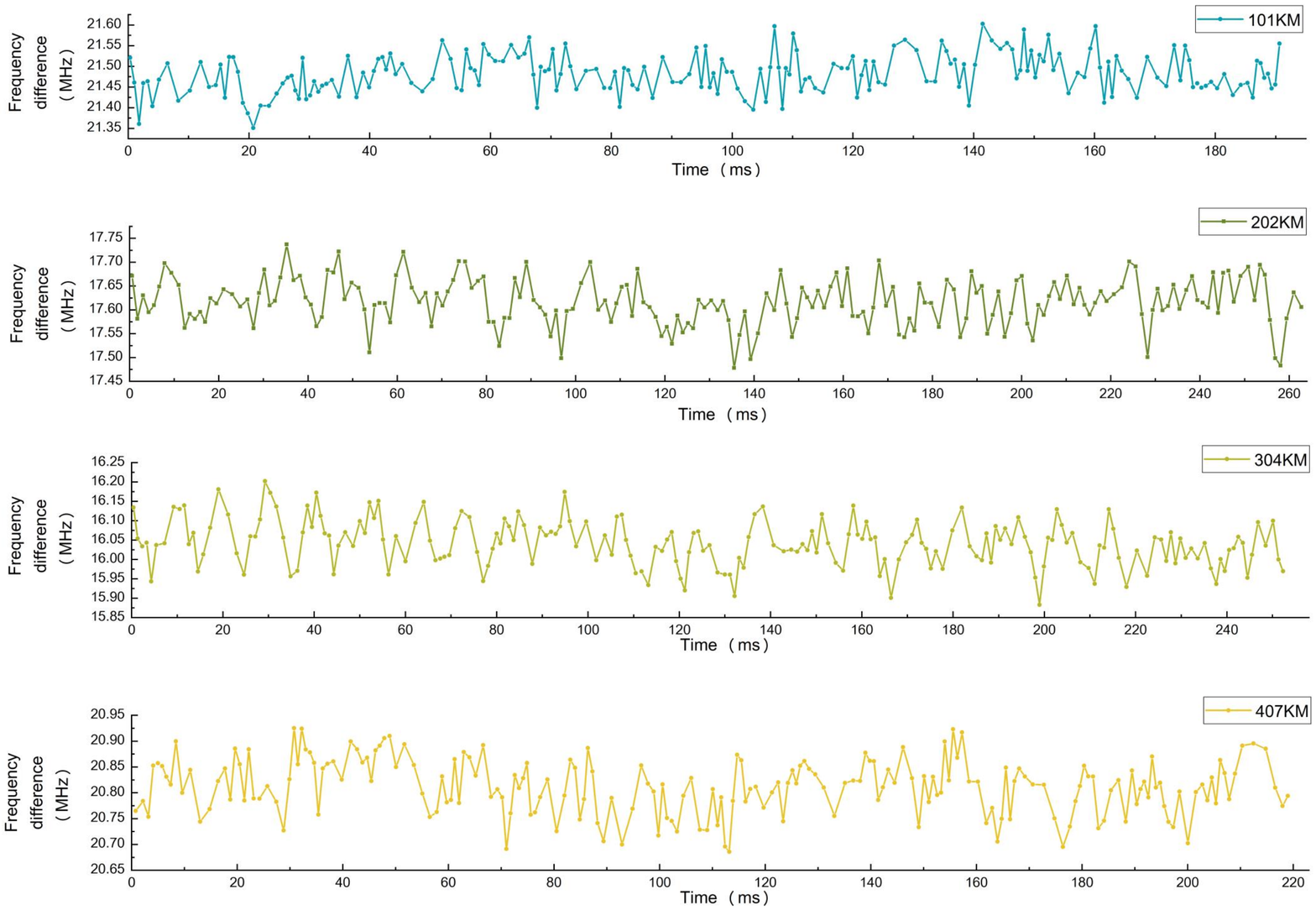}
	\caption{Frequency estimation results. We use the MLE method to estimate the frequency difference under every 101, 202, 304, and 407 km. The results for every distance contains 200 times of estimation. It shows that the fluctuation of the frequency differences is not large.}
	\label{fig:Freqestimationdrift}
\end{figure*}

Using the MLE method mentioned in Sec.~\ref{subsec:MLE}, Alice and Bob estimate the frequency difference for every 500 effective clicks. Here, we choose 500 effective clicks for one estimation because the simulation results show that 500 clicks can give a relatively accurate $\Delta\omega(t)$. Meanwhile, it takes no more than 1 ms to accumulate 500 strong light pulses' clicks so that we can treat $\Delta\omega(t)$ as a constant. Afterwards, Alice and Bob utilize 200 estimated phase references to fit $\Delta\omega(t)$ for QKD pulses by least square estimation. We give some frequency difference estimation results under 101, 202, 304, and 407 km in Fig.~\ref{fig:Freqestimationdrift} to show that the estimated frequency difference does not fluctuate much over time.
%Then, the $X$-error rate of signal pulses can be calculated according to the frequency difference given by $\Delta\omega(t)$.

Intuitively, the more accurate the frequency difference is, the lower $X$-basis error rate one can get. So we can evaluate the accuracy of phase reference estimation by applying the result to the pairs of strong light pulses, similar to the $X$-basis error calculation of QKD pulses. Here, the strong light pulses are not encoded with any phase information. Also, we show the relationship between the error rate of strong pulses and the $X$-basis error rate of QKD under 202 km in Table.~\ref{table:FrequencyEstimationErr}. The results show that the two error rates are well aligned. 

\begin{table}[htbp!]
	\caption{Relationship between the error rates for the pairs of strong light pulses and pairs of QKD pulses under different intervals of pairing length. To get more pairs in the reference region to improve the accuracy of phase reference estimation, one click in reference region may be paired repeatedly. This test carries out under the condition of 202 km fiber.}
	\vspace{20pt}
	\centering
	\begin{tabular}{p{2cm}|p{1.5cm}|p{1.5cm}|p{1.5cm}|p{1.5cm}|p{1.5cm}|p{1.5cm}|p{1.5cm}|p{1.5cm}}
		\hline
		\hline
		\hspace{2em}\multirow{2}*{L} & \multicolumn{4}{c|}{Strong pulses pairs} & \multicolumn{4}{c}{QKD pulses pairs} \\
		\cline{2-9}
		~ &  Total Pairs & Error Pairs & Error Rate & Standard Deviation & Total Pairs & Error Pairs & Error Rate& Standard Deviation\\
		\hline
		\hspace{0em} [63,500) &  27817087 & 7716898 & 0.2774 & 0.0022 & 261923 & 79234 & 0.3025 & 0.0028
		\\
		\hline
		\hspace{0em} [500,1000) &  26194467 & 7409449 & 0.2829 & 0.0027 & 146520
		& 45185 & 0.3084 & 0.0050\\		
		\hline
		\hspace{0em} [1000,1500) &  19972305 & 5871640 & 0.2940 & 0.0048 & 67458 & 21410 & 0.3174 & 0.0083\\	
		\hline
		\hspace{0em} [1500,2000) &  13845880 & 4283679 & 0.3094 & 0.0088 & 30799 & 10150 & 0.3296 & 0.0117\\
		\hline  
		\hline  
	\end{tabular}
	\label{table:FrequencyEstimationErr}
	
\end{table}

\begin{figure}[htbp]
	\includegraphics[width=8cm]{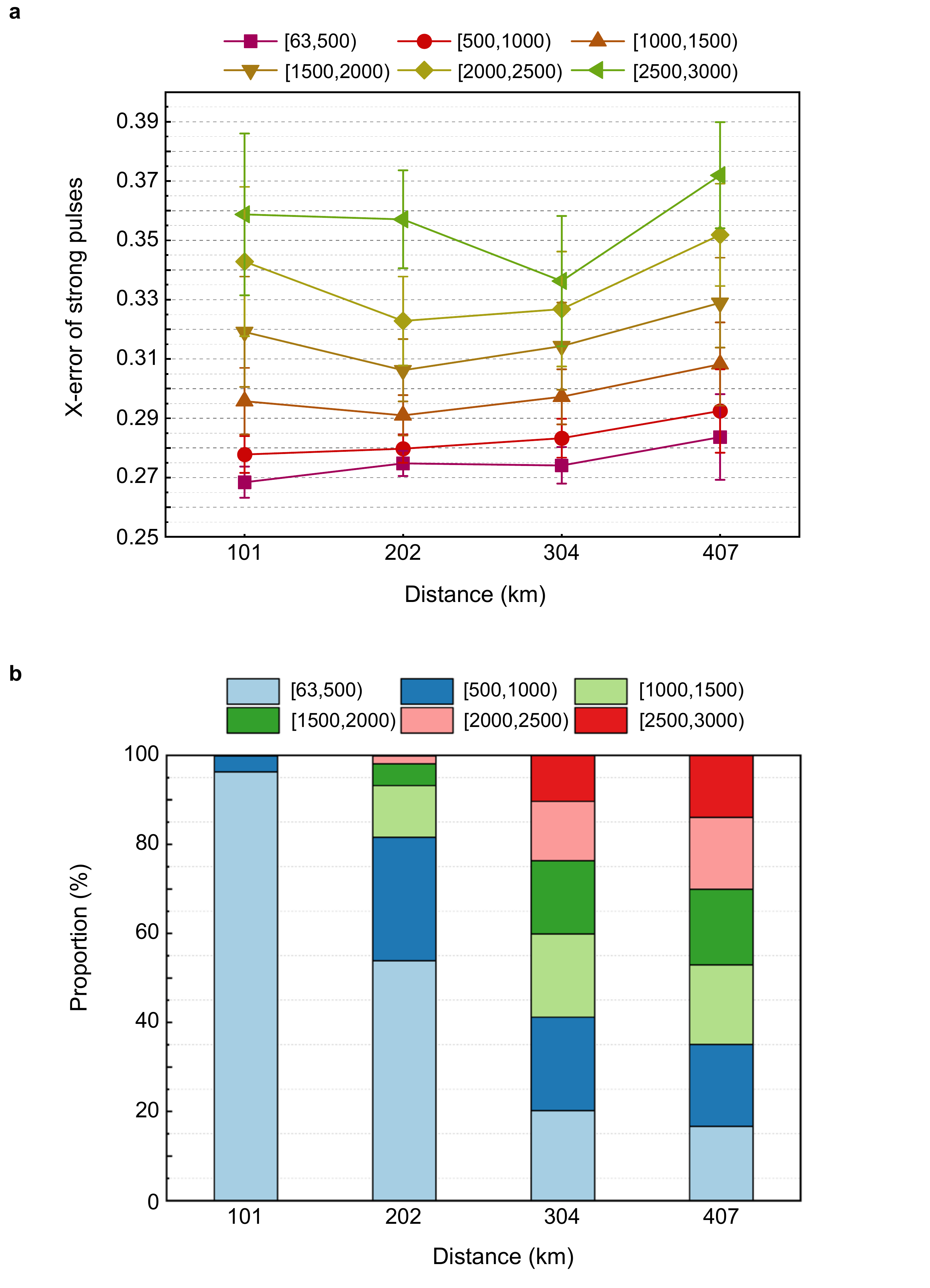}
	\caption{
		$X$-error rates and pairs ratios for different $l$. (a) The $X$-basis error rate of strong light pulses, which can be used to quantify how well Alice and Bob estimate the phase reference. It shows that the $X$-basis error rate also varies under different fiber lengths and pair intervals. A larger pairing length interval results in larger phase fluctuations. Consequently, the $X$-basis error rate increases with the communication distance. The minimum pairing length is chosen to be 63 to avoid the errors caused by afterpulse and dead-time of the detectors. %The number of pulses tested here is larger than $5\times10^{11}$. 
		(b) The ratio of pairs falling into certain pairing length intervals changes with the fiber distances when the maximum pairing length is chosen to be 3000. The interval $[63, 500)$ means that only two clicked pulses with an index difference between 63 and 500 are counted. Other intervals are defined similarly. For a short communication distance, the ratio for low intervals like $[63, 500)$ is high, because the average pair length is small. While for long distances, the ratio for high intervals like $[1500, 2000)$ becomes larger.
	} \label{fig:diffL}
\end{figure}

Then, in order to pick up proper pairing length intervals for different trials, we test the $X$-basis error rate for the strong pulses under different pairing lengths and transmission distances, as shown in Fig.~\ref{fig:diffL}(a). Besides, we also check the ratios of clicked pairs falling into certain pairing length intervals, as shown in Fig.~\ref{fig:diffL}(b). 

In addition to the maximum pairing length $L_{\text{max}}$, we also set the minimum pairing length $L_{\text{min}}=63$ to avoid the influence of the detector dead time and after-pulse effect. So here we need to make some corrections to the pairing rate in \cite{zeng2022quantum}. The new pairing rate considering both maximum and minimum pairing lengths is given by,
\begin{equation}\label{eq:pairR}
	r_p(p,L_{\text{min}},L_{\text{max}})=\left[\dfrac{1}{p\left[(1-p)^{L_{\text{min}}-1}-(1-p)^{L_{\text{max}}}\right]}+\dfrac1p\right]^{-1},
\end{equation}
where $p$ is the probability that the emitted pulses result in a click event and has the scaling of $O(\sqrt{\eta})$. This equation shows that after introducing $L_{\text{min}}$, the pairing rate is roughly inversely proportional to $\sqrt{\eta}$. The average pairing length is proportional to $r_p^{-1}$ because $r_p$ is defined as the average number of pairs generated per pulse. Therefore, when the communication distance is short, the average pairing length will be short and the ratio of pairs falling into short intervals will be high. When the communication distance increases, the average pairing length will be longer and more pairs will fall into longer intervals. In this case, to make more clicks be paired, we need to increase the maximum pairing length and get more pairs.

As shown in Fig.~\ref{fig:diffL}(a), the $X$-basis error rate increases with the pairing length $l$ but does not vary much under different distances. As mentioned in Sec.~\ref{Sc:Phase drift}, the contribution of fiber phase drift is relatively small, so the error rate does not vary much when the fiber length increases. As for the reason why the $X$-basis error rate increases with a larger pairing length, the MLE method we use can only estimate the phase difference in a period of time after accumulating enough strong pulse clicks while the reference is still changing in the period due to the change of two lasers’ frequency. The accumulating time is typically milliseconds due to the limitation of the SNSPD counting rate and strong pulse frequency. The estimation results are the average frequency difference during each accumulation time, which could deviate from the actual value. We further adopt the least square method to improve the estimation accuracy. A longer $l$ will make the integral interval in Eq.~\eqref{eq:MPfreq} larger, further amplifying the error of $\Delta \omega$. Eventually, the estimation results of the phase difference will be less accurate, leading to a higher $X$-basis error rate.
%lead to inaccurate phase estimation, which will increase the $X$-basis error rate. 
Therefore, to achieve higher key rates, we cannot set $L_{\text{max}}$ too big to maintain a reasonable $X$-basis error rate. After considering the trade-off between the pairing number and the $X$-basis error rate, we choose $L_{\text{max}}=500$ for 101 km, $L_{\text{max}}=1000$ for 202 km, and $L_{\text{max}}=2000$ for 304 and 407 km.

%\subsection{Test results of different pairing length}

\section{Experimental details} \label{sc:ExprDetail}
\subsection{Details of experimental setup}
For system synchronization and electrical signals generation, two arbitrary-function generators (Tektronix, AFG3253) provide 100 KHz electrical signals as a trigger source. The synchronization signals are also used for Charlie. As a result, Alice, Bob and Charlie share the trigger source for generating 625 MHz signals and the system clock. The modulators are driven by the 625 MHz electrical signals with a field-programmable gate array. Charlie adjusts the delay of 100 kHz electrical signals to make two light pulses arrive at Charlie’s beam splitter simultaneously.

Additionally, the Sagnac ring contains a circulator, a polarization-maintaining beam splitter (PMBS), a phase modulator, and a fiber. The pulses pass through one port of the PMBS and are divided into two pulses. Then the two pulses pass through the same fiber from different directions, where the phase modulator is allocated, and arrive at another port simultaneously. The fiber lengths two pulses passing through before the phase modulator are different. 
By adjusting this difference, we can design the Sagnac ring for different system repetitions. The two divided pulses pass through the same fiber, the Sagnac ring can be stable and has a high extinction ratio. 

On Charlie’s side, two electric polarization controllers (EPCs) and two polarization beam splitters (PBSs) are allocated for polarization basis alignment. Photons from one side of the PBSs are monitored by two superconducting nanowire single-photon detectors (SNSPDs). Charlie performs polarization feedback by adjusting the voltage loaded in EPC to increase the transitivity of one arm of the PBS. Then, after this feedback, the polarization of light pulses for interference measurement is nearly identical. Besides, the arriving time of photons from the two SNSPDs is recorded for feedback, by adjusting the delay of the electrical signal for time calibration. High-quality SNSPDs are employed for photon detection.

\subsection{System tests} \label{Sc:Laser frequency drift}
\begin{figure}[htbp]
	\includegraphics[width=18cm]{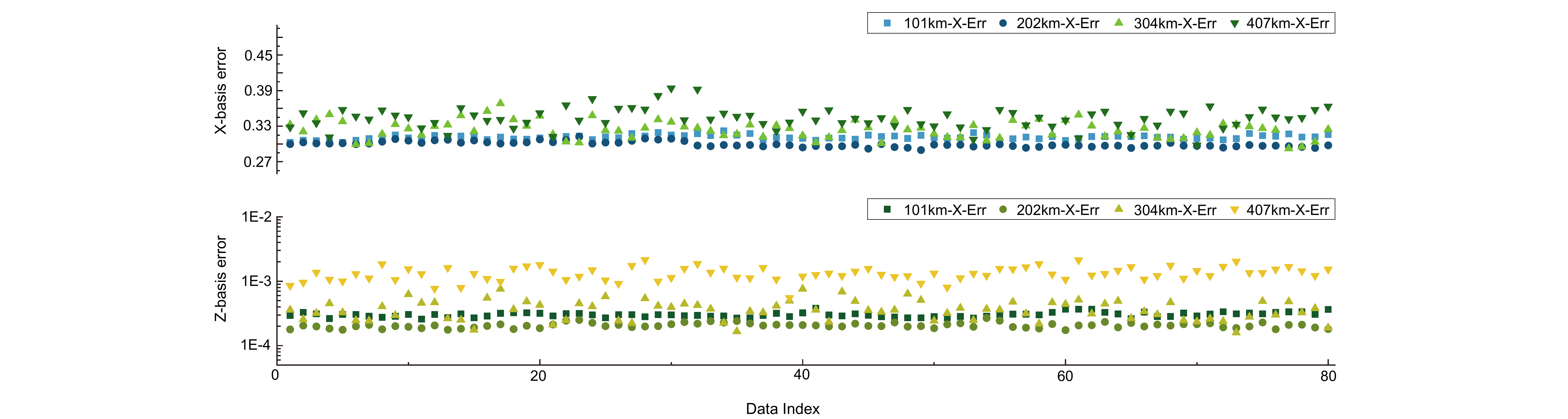}
	\caption{System stability. The $X$-basis and $Z$-basis error rates varied with time under different fiber lengths. For 101 km, each data point represents the effective clicks collected in 10 s. For 202 km and 304 km, each data point represents the effective clicks collected in 30 s. For 407 km ULL fiber, each data point represents the effective clicks collected in 150 s.
		%The X-basis and Z-basis error rates varied with time under different communication distances. For 101 km, each data point represents the effective clicks collected in 10 s. For 202 and 304 km, each data point represents for 30 s. For 407 km, each data point represents for 150 s.
	} \label{fig:otherresults}
\end{figure}

We also the stability of the system by recording how $Z$-basis and $X$-basis error rates vary with time under different fiber lengths. The results are shown in Fig.~\ref{fig:otherresults}, which indicates that $X$-basis error rate varies little under different fiber lengths, which is about $31\%$ to $34\%$. Besides, the $Z$-basis error is also stable and relatively low. These results show that our system is quite stable when running continuously.

\begin{figure}[htbp!]
	\includegraphics[width=10cm]{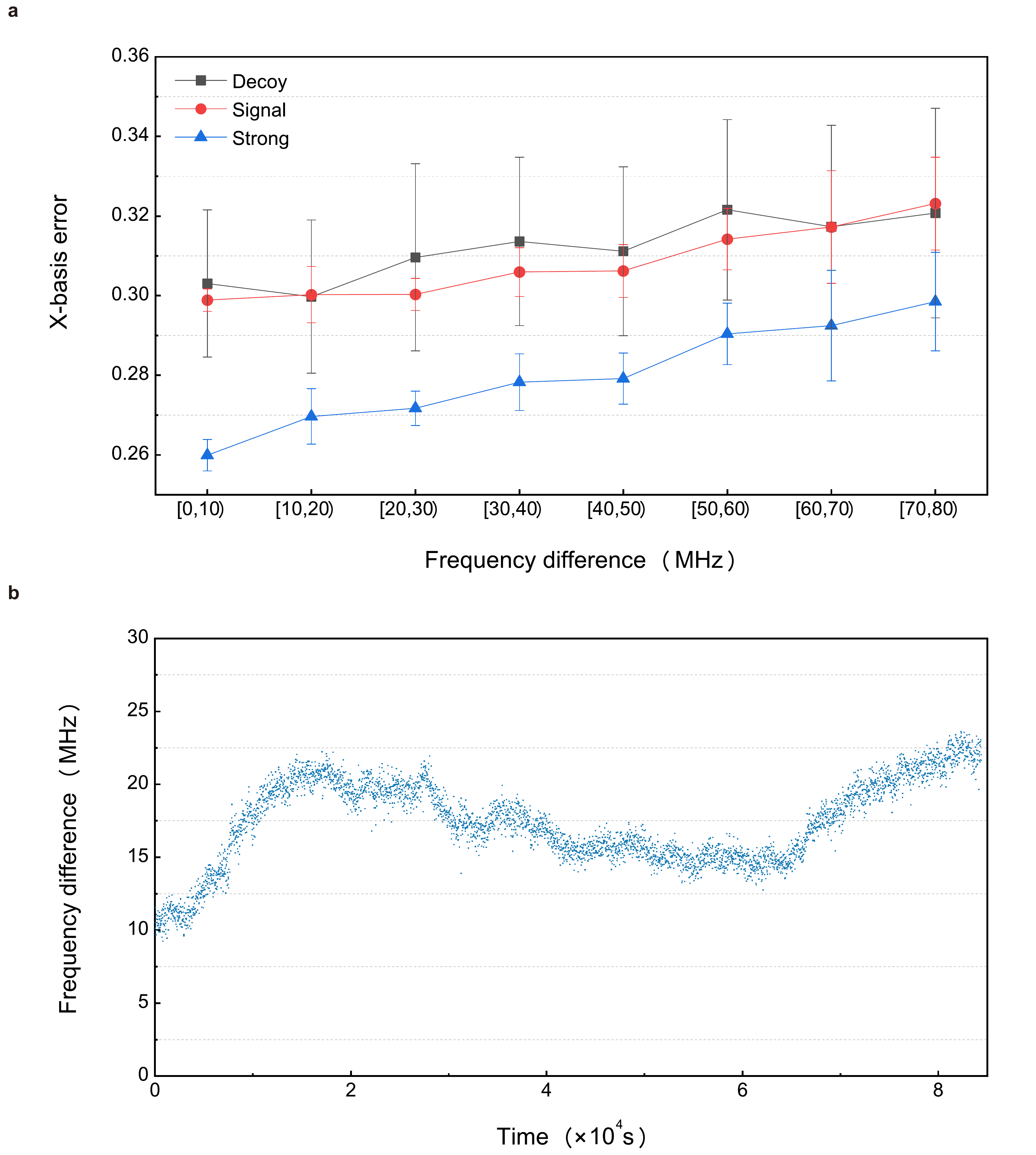}
	\caption{(a) Error rates under different laser frequency differences. Decoy: decoy-state pulses. Signal: signal-state pulses. Strong: strong light pulses. In order to get enough click data to ensure the accuracy of the results, the number of sent pulses of each data point changes from $5\times10^{11}$ to $2\times10^{12}$. The results show that the $X$-basis error rate increases with the increase of frequency difference while the $X$-basis error rate is still relatively low even if the frequency difference reaches 100 Mhz. (b) Frequency differences over time. The interval between two data points is 20 s. The frequency differences between Alice's and Bob's lasers are directly read from a frequency counter (Keysight 53220A). The results show that the frequency differences is relatively stable.}
	\label{fig:XerrorVsFreq}
\end{figure}

Without phase locking, we also need to test what the maximum frequency difference between Alice's and Bob's lasers is allowed in MP QKD and decide how to carry out the laser frequency feedback. We take measurements for different frequency differences under 202 km fiber and show the results in Fig.~\ref{fig:XerrorVsFreq}(a). One can see that MP QKD can work well even when the frequency difference is as large as 100 MHz. Hence, there is no need of sophisticated laser frequency feedback. In practice, we only need to adjust the wavelength of laser on one side after a long period, even with off-the-shelf lasers. Also, we give the frequency difference drift in several hours by the frequency counter without feedback in Fig.~\ref{fig:XerrorVsFreq}(b).

\section{More comparison and discussion}\label{sc:discussion}
In the main text, we compare the key rate performance of our system with the linear key rate bound. In this section, we will compare the MP and one-mode schemes. Here, we take phase-matching (PM) scheme \cite{Ma2018phase} as an example of the one-mode scheme.

\begin{figure}[htbp!]
	\includegraphics[width=10cm]{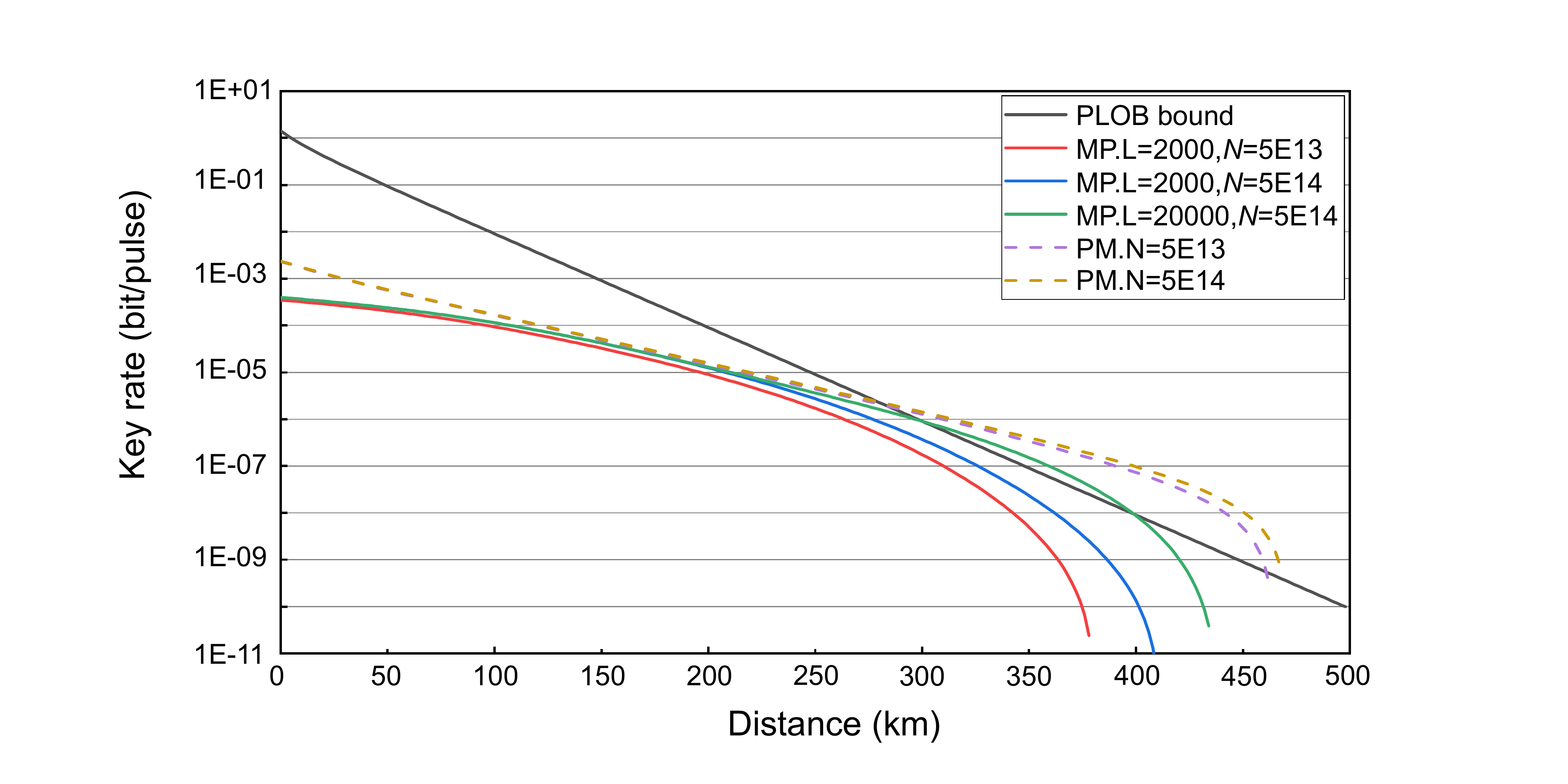}
	\caption{Performance of the PM and MP protocols. The simulation is based on parameters of our current system. The fiber loss is set to be $0.2 \text{ dB/km}$. The detector dark count is $2.72\times 10^{-8}\text{/pulse}$ and the detection efficiency is $62.46\%$. The total number of rounds is denoted as $N$. The misalignment error in the PM scheme is set to be $5.3\%$, based on \cite{fang2020implementation}. The error rate of $X$-pairs in the MP scheme uses $33\%$ of the $304\text{ km}$ experimental results.}
	\label{fig:PMcompare}
\end{figure}

In Fig.~\ref{fig:PMcompare}, we simulate the key rates of the MP scheme and the PM QKD scheme under different parameters. The simulation method of the PM scheme and some parameters such as $e_d$ come from Ref.~\cite{fang2020implementation}. We unify the unit of key rate into bits per pulse for comparison, which will make the key rate value of the MP scheme become half of that in Fig.~3 of the main text. We need to emphasize again that the key rate of one-mode schemes is zero under the same setting as the MP scheme since the phase differences between Alice and Bob are almost random without global phase locking. The key rate of the PM scheme is simulated with an additional assumption that Alice and Bob carry out precise global phase locking.
%We need to emphasize again that this comparison is unfair in practice. The key rate of one-mode schemes is zero under the same setting as the MP scheme since the phase differences between Alice and Bob are almost random without global phase locking.
%give more simulation results and discuss the differences between the two schemes.
% We also discuss the possibility of breaking the linear bound and the reasons for the differences between MP QKD and other protocols.

As we can see in the figure, the key rate scaling of the PM scheme is maintained at $O(\sqrt{\eta})$ before about 400 km. When the communication distance is longer than 400km, the influence of factors such as the detector dark count becomes significant, making the key rate decline rapidly. 

For the MP scheme, the key-rate scaling is more complicated. We need to analyze dominant factors for different communication distances. At short distances, the interval between the clicks is small because the click probability is high. Here, we need to set the minimum pair length $L_\text{min}=63$ due to the detector dead time and the after-pulse effect. So, many clicks in this case are discarded because they are too close to other clicks. This makes the key rate of the MP scheme lower than that of the PM scheme at short distances. When the distances become longer, the interval will be larger and more clicks can be paired to generate key bits. The effect of the minimum pair length $L_\text{min}=63$ becomes smaller. When the communication distance is about 100km-200 km, almost all the clicks will be paired and the number of pairs will be saturated. The scaling of key rate in this case is about $O(\sqrt{\eta})$, which is consistent with the ideal case analyzed in the theoretical paper \cite{zeng2022quantum}. The key rate performance in the figure also shows that the slopes of the PM and MP curves are the same and the key rate values of the two schemes are comparable at these communication distances. Here, because the PM scheme only needs one pulse to generate the raw key bits, while the MP scheme needs a pair of pulses, the key rate of the PM scheme is roughly twice as high as that of the MP scheme.

%When the communication distance exceeds 200 km, the main factor affecting the bit rate will become the maximum pairing length.
%The results show that the PM scheme is more sensitive to the data size. Because the intensity of pulses prepared by Alice and Bob in PM scheme is relatively low, typically $10^{-2}$, the number of clicks in PM scheme is lower. Thus, when the total rounds is small ($N=6\times10^{12}$), the key rate of PM scheme is greatly affected by the finite-size effect and cannot surpass the linear bound. For a larger data size, $N=5\times 10^{13}$, the key rate of PM scheme is improved at all communication distances, and it can surpass the linear bound. As for the MP scheme, since the intensity of signal state is typically $10^{-1}$, the key rate is less affected by the data size, especially at short communication distances. 
At short distances (0-200 km), the maximum pairing length $L_{\text{max}}$ has little effect on the key because $L_{\text{max}}=2000$ is large enough to saturate the number of pairs. Almost all the valid clicks will be paired to generate key bits or do parameter estimation. At longer communication distances (200-400 km), because the $L_{\text{max}}$ that choose in this part is not large enough to make all the clicks to be paired. As $L_{\text{max}}$ expands, more responses can be paired to generate key bits so that the final key rate will also increase. But as mentioned before, longer pairs have higher $X$-basis error rates. So, we cannot set a too large $L_{\text{max}}$, or the overall $X$-basis error rate will be too large to generate any final key bits. 

Intuitively, the product of the click probability $P_{\text{click}}$ and the maximum pairing length $L_{\max}$ tells us how many clicks are expected to be present within $L_{\max}$ pulses after one click. So we can use this value to analyze the pairing roughly. We give the experimental results of $P_{\text{click}}$ and its product with $L_{\text{max}}$ in Table \ref{table:ptimel}. If the product value is larger than 1, there is likely other clicks within $L_\text{max}$ after a given click. In this case, most of the clicks could be paired, corresponding to the experiment's 101 km and 202 km cases. If we can pair most of the click results, the pairing rate is roughly equal to the probability of a single click, i.e., $O(\sqrt{\eta})$. So, in this case, the scaling of the final key rate will be $O(\sqrt{\eta})$, which is consistent with our conclusion that the key rate-transmittance relation of our system follows $R=O(\sqrt{\eta})$ under the intercity communication distances. For longer distances (304 km and 407 km), the product value is one order of magnitude less than 1, so there will be many clicks that cannot be paired, making the key-rate scaling below $O(\sqrt{\eta})$. 
%That is why our experimental results cannot surpass the linear bound. 

\begin{table}[bpth!]
	%\vspace{10pt}
	\centering
	\caption{Experimental results of click probability and their product with $L_{\text{max}}$. }
	\label{table:ptimel}
	\begin{tabular}{l|llll}
		\hline
		\hline
		& $101$ km  & $202$ km & $304$ km & $407$ km \\  
		\hline		
		$\textit{P}_{\text{click}}$& $6.35\times10^{-3}$ & $1.13\times10^{-3}$ & $1.89\times10^{-4}$ & $5.00\times10^{-5}$ \\				
		$L_{\max}$& 500& 1000 & 2000 & 2000 \\
		$L_{\max}\times\textit{P}_{click}$& 3.18 & 1.13 & 0.378 & 0.100\\
		\hline    
		\hline    	
	\end{tabular}
\end{table}

In addition, the data size also influences the key rate of the MP scheme. This is because we need to use $X$-pairs for parameter estimation. Considering the finite-size effect, when the number of total rounds is small, Alice and Bob need to send more decoy states with intensity $\nu$ to ensure a sufficient number of $X$-pairs for parameter estimation. Then, the ratio of rounds sending signal states with intensity $\mu$ will be smaller, decreasing the number of $Z$-pairs for key distillation. On the other hand, when the total round number is larger, the ratio of signal states sent by Alice and Bob can be increased, resulting in more $Z$-pairs for generating raw key bits. Therefore, the simulation curves of the MP scheme shows that larger data sizes would yield higher key rates in Fig.~\ref{fig:PMcompare}. In the asymptotic case, the sending ratio of decoy states is close to zero, so the ratio of signal states will be larger which increases the number of $Z$-pairs. Therefore, the key rate in the asymptotic case will be significantly larger than that in finite-data-size cases, as shown in Fig.~3 of the main text.
%Considering the signal states are paired to form $Z$-pairs, 

If we increase the system frequency by a factor of 10, we can increase $L_\text{max}$ by a factor of 10 while maintaining the $X$-basis error rate. In this case, the product of $P_{\text{click}}$ and $L_{\max}$ will be more than 1 in long communication distances so that the final key rate can surpass the linear bound. We also give the simulation curve with $N=5\times 10^{14}$ in Fig.~\ref{fig:PMcompare}, proving that with $L_{\max}= 20000$ our system is able to surpass the linear bound. Note that increasing the frequency of the lasers is not a hard limitation in our system. In fact, the frequency is mainly limited by our design of Sagnac rings and electronic controls.

\section{Experimental results and raw data}\label{sc:Exprdata}

\begin{table*}[]
	\vspace{10pt}
	\centering
	\setlength{\abovecaptionskip}{0cm} 
	\setlength{\belowcaptionskip}{0.5cm}	
	\caption{Experimental results. $\eta$ is the channel transmittance of single link and $\textit{P}_{click}$ represents single-round click probability. $\textit{r}_{p}$ represents single-round pairing probability. Note that the number of effective QKD pulses sent per second is 444.83 M. Among the rest pulses, 160.97 M strong light pulses are used for phase reference estimation and 19.20 M recovery pulses are vacuum. 
	}
	
	\begin{tabular}{p{3cm}|p{3cm}p{3cm}p{3cm}p{3cm}}
		\hline
		\hline
		Distance(km) & $101$  & ${202}$ & ${304}$ & ${407}$  \\  
		\hline
		$\alpha$ & $ 0.208 $ & $ 0.181 $ & $ 0.186 $  & $ 0.162 $ \\
		$\eta$ &$4.32\times10^{-2}$&$6.80\times10^{-3}$&$7.43\times10^{-4}$&$2.18\times10^{-4}$\\
		$\mu$ & $0.309$ & $0.338$ & $0.531$ & $0.429$ \\
		$\nu$ & $0.032$ & $0.035$ & $0.053$ & $0.038$ \\
		$\textit{p}_\mu$ & $0.22$ & $0.25$ & $0.20$ & $0.23$ \\
		$\textit{p}_\nu$ & $0.18$ & $0.25$ & $0.40$ & $0.41$ \\
		$\textit{N}$ & $ 5.07\times10^{11} $ & $ 2.10\times10^{12} $ & $ 6.33\times10^{12} $  & $ 7.66\times10^{13} $ \\	
		$L_{max}$& 500& 1000 & 2000 & 2000 \\	
		\hline
		$\textit{e}^{ph}_{11}$& 0.2474 & 0.2571 & 0.3370 & 0.3468 \\		
		$\textit{P}_{click}$& $6.35\times10^{-3}$ & $1.13\times10^{-3}$ & $1.89\times10^{-4}$ & $5.00\times10^{-5}$ \\				
		$\textit{r}_{p}$& $2.46\times10^{-3}$ & $4.29\times10^{-4}$ & $4.42\times10^{-5}$ & $4.21\times10^{-6}$ \\			
		$\textit{M}^{Z,L}_{11}$ & $1.07\times10^{8}$ & $5.03\times10^{7}$ & $4.51\times10^{6}$ & $6.02\times10^{6}$ \\			
		$\textit{QBER}$ & $3.10\times10^{-4}$ & $1.89\times10^{-4}$ & $3.75\times10^{-4}$ & $1.46\times10^{-3}$ \\	
		\hline
		Key bits per pair& $7.75\times10^{-5}  $& $9.34\times10^{-6}$&$8.65\times10^{-8}$&$3.46\times10^{-9}$ \\
		key bits per second &$1.72\times10^{4} $ &$2.08\times10^{3}$&$1.92\times10^{1} $& $7.69\times10^{-1} $\\		
		\hline
		$Sent-A_0B_0$ & $182558232000$ & $523787325000$ & $1014212400000$ & $9929053988640$ \\		
		$Sent-A_0B_\nu$ & $54767469600$ & $261893662500$ & $1014212400000$ & $11308089264840$ \\
		$Sent-A_0B_\mu$ & $66938018400$ & $261893662500$ & $507106200000$ & $6343562270520$ \\
		$Sent-A_\nu B_0$ & $54767469600$ & $261893662500$ & $1014212400000$ & $11308089264840$ \\
		$Sent-A_\nu B_\nu$ & $16430240880$ & $130946831250$ & $1014212400000$ & $12878657218290$ \\						
		$Sent-A_\nu B_\mu$ & $20081405520$ & $130946831250$ & $507106200000$ & $7224612585870$ \\		
		$Sent-A_\mu B_0$ & $66938018400$ & $261893662500$ & $507106200000$ & $6343562270520$ \\		
		$Sent-A_\mu B_\nu$ & $20081405520$ & $130946831250$ & $507106200000$ & $7224612585870$ \\		
		$Sent-A_\mu B_\mu$ & $24543940080$ & $130946831250$ & $253553100000$ & $4052831450610$ \\
		\hline		
		$Detected-A_0B_0$ & $371839$ & $113728$ & $82262$ & $2057817$ \\		
		$Detected-A_0B_\nu$ & $70909222$ & $59920247$ & $40834074$ & $96106357$ \\
		$Detected-A_0B_\mu$ & $852483360$ & $579742840$ & $203014904$ & $612172471$ \\
		$Detected-A_\nu B_0$ & $72327873$ & $64181636$ & $39195652$ & $98605251$ \\
		$Detected-A_\nu B_\nu$ & $42910163$ & $62048474$ & $79923680$ & $219103424$ \\						
		$Detected-A_\nu B_\mu$ & $281553032$ & $321774876$ & $222609546$ & $758824857$ \\		
		$Detected-A_\mu B_0$ & $844185844$ & $627681721$ & $196494573$ & $609953543$ \\		
		$Detected-A_\mu B_\nu$ & $278465266$ & $343488003$ & $216791057$ & $754579095$ \\		
		$Detected-A_\mu B_\mu$ & $614776758$ & $602577240$ & $199695473$ & $780414728$ \\

		\hline    
		\hline    
		
	\end{tabular}
	
	\label{table:ExpResult_SingleRound}
	
\end{table*}

%- 200,300 km data

\begin{table*}[]
	
	\vspace{10pt}
	\centering
	
	\setlength{\abovecaptionskip}{0.5cm} 
	\setlength{\belowcaptionskip}{0.5cm}	
	\caption{Number of pairs for different cases. \textit{Total}: the total pair number of all clicks. \textit{Error}: the number of error pairs. Especially, 0 means Alice or Bob sends two vacuum state for two clicks of the pair. $ \nu $ means Alice or Bob sends one decoy state and one vacuum state. $ \mu $ means Alice or Bob sends one signal state and one vacuum state. 2$ \nu $ means Alice or Bob sends two decoy states. 2$ \mu $ means Alice or Bob sends two signal states.}
	\begin{tabular}{p{3cm}|p{2.5cm}p{2.5cm}p{2.5cm}p{2.5cm}}
		\hline
		\hline
		Distance & $\rightline{101}$  & $\rightline{202}$ & $\rightline{303}$ & $\rightline{407}$  \\  
		\hline
		
		$Z_{A0B0}-Total$ & $\rightline{395}$ & $\rightline{1}$ & $\rightline{3}$ & $\rightline{79}$ \\
		
		$Z_{A0B0}-Error$ & $\rightline{395}$ & $\rightline{1}$ & $\rightline{3}$ & $\rightline{79}$ \\
		
		$Z_{A0B\nu}-Total$ & $\rightline{7624}$ & $\rightline{2089}$ & $\rightline{1130}$ & $\rightline{3235}$ \\
		
		$Z_{A0B\nu}-Error$ & $ \rightline{3799} $ & $ \rightline{1073} $ & $ \rightline{594} $  & $ \rightline{1615} $ \\
		
		$Z_{A0B\mu}-Total$ & $ \rightline{89296} $ & $ \rightline{19612} $ & $ \rightline{5674} $  & $ \rightline{18641} $ \\
		
		$Z_{A0B\mu}-Error $ & $ \rightline{43755} $ & $ \rightline{9882} $ & $ \rightline{2860} $  & $ \rightline{9379} $ \\
		
		$Z_{A\nu B0}-Total$ & $ \rightline{7796} $ & $ \rightline{2170} $ & $ \rightline{1183} $  & $ \rightline{3088} $ \\
		
		$Z_{A\nu B0}-Error$ & $ \rightline{3833} $ & $ \rightline{1067} $ & $ \rightline{586} $  & $ \rightline{1525} $ \\
		
		$Z_{A\nu B\nu}-Total$ & $ \rightline{1484349} $ & $ \rightline{1151368} $ & $ \rightline{589395} $  & $ \rightline{401910} $ \\
		
		$Z_{A\nu B\nu}-Error$ & $ \rightline{4510}  $ & $ \rightline{2095} $ & $ \rightline{2281} $  & $ \rightline{6839} $ \\
		
		$Z_{A\nu B\mu}-Total$ & $ \rightline{17793988}  $ & $ \rightline{11133038} $ & $ \rightline{2890760} $  & $ \rightline{2593713} $ \\
		
		$Z_{A\nu B\mu}-Error$ & $ \rightline{29614}  $ & $ \rightline{10871} $ & $ \rightline{6122} $  & $ \rightline{22878} $ \\	
		
		$Z_{A\mu B0}-Total$ & $ \rightline{87784}  $ & $ \rightline{21241} $ & $ \rightline{5491} $  & $ \rightline{18580} $ \\
		
		$Z_{A\mu B0}-Error$ & $ \rightline{42577}  $ & $ \rightline{10530} $ & $ \rightline{2770} $  & $ \rightline{9431} $ \\		
		
		$Z_{A\mu B\nu}-Total$ & $ \rightline{17282569}  $ & $ \rightline{11257634}$ & $ \rightline{2951821} $  & $ \rightline{2495312} $ \\
		
		$Z_{A\mu B\nu}-Error$ & $ \rightline{29328}  $ & $ \rightline{11505} $ & $ \rightline{5957} $  & $ \rightline{22525} $ \\		
		
		$Z_{A\mu B\mu}-Total$ & $ \rightline{207389568}  $ & $ \rightline{108781949} $ & $ \rightline{14657476} $  & $ \rightline{16100111} $ \\
		
		$Z_{A\mu B\mu}-Error$ & $ \rightline{64238}  $ & $ \rightline{20550} $ & $ \rightline{5494} $  & $ \rightline{23545} $ \\

		$X_{A0 B2\nu}-Total$ & $ \rightline{728841}  $ & $ \rightline{540202} $ & $ \rightline{312052} $  & $ \rightline{195135} $ \\
		
		$X_{A0 B2\nu}-Error$ & $ \rightline{371608}  $ & $ \rightline{269911} $ & $ \rightline{155379} $  & $ \rightline{96467} $ \\		
		
		$X_{A0 B2\mu}-Total$ & $ \rightline{104986314}  $ & $ \rightline{50434095} $ & $ \rightline{7705544} $  & $ \rightline{8211680} $ \\
		
		$X_{A0 B2\mu}-Error$ & $ \rightline{53565167}  $ & $ \rightline{25204023} $ & $ \rightline{3840896} $  & $ \rightline{4050328} $ \\		
		
		$X_{A2\nu B0}-Total$ & $ \rightline{757883}  $ & $ \rightline{620721} $ & $ \rightline{285196} $  & $ \rightline{213627} $ \\
		
		$X_{A2\nu B0}-Error$ & $ \rightline{387259}  $ & $ \rightline{310412} $ & $ \rightline{142195} $  & $ \rightline{105084}$ \\

		$X_{A2\mu B0}-Total$ & $ \rightline{102882407}  $ & $ \rightline{59256669} $ & $ \rightline{7160291} $  & $ \rightline{8244022} $ \\
		
		$X_{A2\mu B0}-Error$ & $ \rightline{52564215}  $ & $ \rightline{29637863} $ & $ \rightline{3573229} $  & $ \rightline{4063596} $ \\

		$X_{A2\nu B2\nu}-Total$ & $ \rightline{33604}  $ & $ \rightline{72245} $ & $ \rightline{147652} $  & $ \rightline{129574} $ \\
		
		$X_{A2\nu B2\nu}-Error$ & $ \rightline{10596}  $ & $ \rightline{22978} $ & $ \rightline{48961} $  & $ \rightline{45226} $ \\		
		
		$X_{A2\nu B2\mu}-Total$ & $ \rightline{1424753}  $ & $ \rightline{1941081} $ & $ \rightline{1153426} $  & $ \rightline{1576219} $ \\
		
		$X_{A2\nu B2\mu}-Error$ & $ \rightline{623114}  $ & $ \rightline{841632} $ & $ \rightline{515419} $  & $ \rightline{714854} $ \\	
		
		$X_{A2\mu B2\nu}-Total$ & $ \rightline{1397813}  $ & $ \rightline{2214830} $ & $ \rightline{1087743} $  & $ \rightline{1565764}$ \\
		
		$X_{A2\mu B2\nu}-Error$ & $ \rightline{611618}  $ & $ \rightline{977362} $ & $ \rightline{480065} $  & $ \rightline{711299} $ \\		
		
		$X_{A2\mu B2\mu}-Total$ & $ \rightline{6811157}  $ & $ \rightline{6799900} $ & $ \rightline{920505} $  & $ \rightline{1658149} $ \\
		
		$X_{A2\mu B2\mu}-Error$ & $ \rightline{2113807}  $ & $ \rightline{2129516} $ & $ \rightline{303172} $  & $ \rightline{566112} $ \\

		\hline    
		\hline    
		
	\end{tabular}

	\label{table:ExpResult_SM}
	
\end{table*}

\clearpage

%\normalem
\bibliographystyle{apsrev4-2}
\bibliography{bibMPExp}

\end{document}